\documentclass[12pt,onecolumn]{IEEEtran}
\usepackage{float}
\usepackage[mathscr]{eucal}
\usepackage{ifpdf}
\usepackage{cite}
\usepackage{bbding}
\usepackage{graphicx}
\usepackage[cmex10]{amsmath}
\usepackage{amssymb}
\usepackage{pifont}
\newcommand{\cmark}{\ding{51}}
\newcommand{\xmark}{\ding{55}}
\usepackage{algorithmic}
\usepackage{units}
\usepackage{setspace}
\usepackage{algorithm}
\usepackage{array}
\usepackage[tight,footnotesize]{subfigure}
\usepackage{amsthm}
\usepackage{slashbox}
\usepackage{multirow}
\usepackage{enumerate}
\usepackage{color}
\usepackage{mathtools}
\newcommand{\subparagraph}{}
\usepackage[compact]{titlesec}
\usepackage{float}
\newtheorem{theorem}{Theorem}
\newtheorem{lemma}{Lemma}
\newtheorem{proposition}{Proposition}

\newtheorem{assumption}{Assumption}

\newtheorem{definition}{Definition}

\newcommand*{\myprime}{^{\prime}\mkern-1.2mu}
\newcommand*{\mydprime}{^{\prime\prime}\mkern-1.2mu}
\begin{document}
\title{Distributed Task Management in Cyber-Physical Systems: How to Cooperate under Uncertainty?}
\author{
\IEEEauthorblockN{Setareh Maghsudi and Mihaela van der Schaar\\}
\thanks{Setareh Maghsudi is with the Department of Electrical Engineering and Computer Science, Technical University of Berlin, 10623 Berlin, Germany (e-mail: maghsudi@tu-berlin.de). Mihaela van der Schaar is with the Department of Engineering Science, University of Oxford, Oxford OX1 2JD, United Kingdam. A part of this paper appeared in IEEE Global Communications Conference 2018.}
}
\maketitle
\begin{abstract}
We consider the problem of task allocation in a network of cyber-physical systems (CPSs). The network can have different states, and the tasks are of different types. The task arrival is stochastic and state-dependent. Every CPS is capable of performing each type of task with some specific state-dependent efficiency. The CPSs have to agree on task allocation prior to knowing about the realized network's state and/or the arrived tasks. We model the problem as a multi-state stochastic cooperative game with state uncertainty. We then use the concept of deterministic equivalence and sequential core to solve the problem. We establish the non-emptiness of the strong sequential core in our designed task allocation game and investigate its characteristics including uniqueness and optimality. Moreover, we prove that in the task allocation game, the strong sequential core is equivalent to Walrasian equilibrium under state uncertainty; consequently, it can be implemented by using the Walras' tatonnement process.  
\end{abstract}
\section{Introduction}
\label{sec:Intro}
With the emergence of complex systems designed to perform a wide variety of tasks, in recent years there has been a significant growth in research efforts on developing efficient task allocation schemes, which find application in a large body of real-world scenarios. This includes domains like web services, smart grid, internet of things, cyber-physical systems, and cloud computing, among many others. In the following section, we briefly review the state-of-the-art research. 
\subsection{State-of-the-Art}
\label{subsec:SoA}
Currently, a body of research works focuses on \textit{heuristic} task management. For example, community-aware task allocation for networked multi-agent systems is considered in \cite{Wang14:CTA} and \cite{deWeerdt12:MTA}, where each agent can negotiate only with its intra-community member agents. Heuristic distributed algorithms are developed to solve the problem. The aforementioned model stands in contrast with global-aware task allocation, where every agent communicates with all other agents in the network. In \cite{Zhao16:AHD}, a heuristic distributed task allocation method is designed for multi-vehicle multi-task problems, where each machine is able to perform multiple tasks, against the single-vehicle models. The authors also investigate the application of the developed solution in search and rescue scenarios. The authors of \cite{Jiang09:CRN} investigate task allocation and load-balancing in complex systems. Here the tolerable load of every network entity depends not only on its own resources, but also on the contextual resources that can be acquired via negotiations. In \cite{Blythe05:TSS}, the authors study the task management problem in a grid. They propose two allocation schemes: (i) Task-based algorithms, that greedily allocate tasks to resources, and (ii) Workflow-based algorithms, that search for an efficient allocation for the entire workflow. Moreover, \textit{Markovian methods} are also applied to model and solve the task allocation problem. As an example, in \cite{Hu17:OTA}, task allocation for human–machine collaborative systems is studied. By modeling the human fatigue as a continuous-time Markov decision process, they show that the optimal task assignment can be solved by linear programming. Article \cite{Dai07:ORA} investigates reliability-based task proportioning and resource allocation. For solving the formulated problem, the authors suggest an algorithm based on graph theory, Bayesian approach, and the evolutionary optimization approach. 

Similar to many other resource management and cost-sharing problems, models adapted from economic theory are widely utilized to address the task allocation and job scheduling problems. In \cite{Ye16:PBSW}, the authors investigate repeated task allocation based on \textit{prospect theory}. They study the effect of the history of task allocation on future agents participation, and the influence of agents' participation on long-term system performance. Location-dependent task allocation for crowd-sourcing is considered in \cite{He17:LAT}. Therein, the authors formulate task allocation as an orienteering problem. In addition, \textit{auction theory} (both single-item and bundle) and mechanism design are among popular theoretical tools to address the task allocation problem. For instance, self-adaptive auction is the basis of task-bundle allocation in \cite{Zhao09:EGT}. Reference \cite{Fu07:TAN} uses a mechanism design approach to address a similar problem. Similar to the auction theory, game theory has been utilized to design distributed task allocation methods. Reference 
\cite{Semasinghe17:GTM} discusses a series of game-theoretical models to address resource management problems in IoT networks, including task allocation. Pilloni \textit{et al.} \cite{Pilloni14:TAN} formulate the distributed task assignment problem as a non-cooperative game, where neighboring nodes engage in negotiations to maximize their own utility scores, resulting in some task allocation. Reference 
\cite{Park09:SCG} addresses the problem by using a model based on Stackelberg convention game model. Distributed allocation of complex tasks in social networks is also considered in \cite{Wang15:MAC}. Two methods, for cooperative and non-cooperative agents, are developed. Task allocation using \textit{coalition formation} has been investigated by some research works so far. Manisterski \textit{et al.} 
\cite{Manisterski06:FEA} propose a centralized task allocation scheme based on cooperation among coalitions of agents, which are built by utilizing minimum-weighted perfect matching. Similarly, in \cite{Kraus03:CFU}, tasks are executed by coalitions of agents. Every agent decides to join a coalition or not through a decentralized auction process. Distributed algorithms for task allocation via coalition formation are studied also in \cite{Shehory98:MTA}. In this work, every task is assigned to one coalition. In addition to the conventional model in which each agent can belong to one coalition only, the authors also study the case in which an agent might belong to multiple coalitions (overlapping coalition formation). Furthermore, \cite{Abdallah06:LTA} formulates the task allocation problem as a 
\textit{repeated game}. The authors develop a distributed mechanism, which uses a mediator to allocate tasks to agents based on a gradient ascent learning algorithm. The goal is to minimize the average turn-around time. In \cite{Grosu02:LBD}, the authors cast the task allocation problem as a cooperative game and apply the Nash bargaining solution to solve the formulated problem. The topic of 
\cite{Birje12:RPS} is resource pricing for wireless grid computing. This paper, models a resource pricing strategy using a non-cooperative bargaining game for resource allocation considering dynamics in the grid market. Reference \cite{Liu05:ABL} investigates the load balancing problem in homogenous mini-grids. The agent-based load balancing is regarded as agent distribution, and two quantities are studied: the number and the size of teams, where agents (tasks) queue. Based on a macroscopic modeling, the load balancing mechanism is characterized using differential equations. By solving the equation, the authors show that the load balancing always converges to a steady state.  
\subsection{Our Contribution}
\label{subsec:Contribution}
In the emerging complex networks, autonomous entities are often required to perform tasks of different types. The tasks to be performed vary over a wide range, including transmission, sensing, measurement, signal processing, computation, and the like. While the type and the load of the jobs arrived at every CPS is in general random, in an inhomogeneous network each CPS might be inclined towards some specific types of tasks, as a consequence of its own characteristics. In other words, non-identical CPSs do not have the same task preferences, or might not be able to perform some randomly-arrived jobs in full due to a lack of resources. 

Fortunately, in the presence of ubiquitous connectivity, the jobs can be performed cooperatively despite the initial random arrival. In order to become promoted, such cooperation shall be beneficial not only at the individual systems' level, but also from the platform's perspective (i.e., with respect to the social welfare). Coordinating such cooperation is however challenging, since often an agreement shall be achieved prior to job arrivals. Moreover, the characteristics of each CPS might depend on the global network's state and vary over time. In general, the state is also unknown at the time of allocation. 

In this paper, we investigate the distributed task management problem in a network of cyber-physical systems, where the stochastic task arrival, as well as the systems' characteristics, are contingent on a random state that is a priori unknown. We model the problem as a stochastic cooperative game with state uncertainty. We first eliminate the randomness in the task arrival by using a notion of expectation, referred to as the \textit{certainty equivalence}. In the resulted deterministic cooperative game, we establish the existence of a special type of core, which is self-enforcing and resistant to the uncertainty in the state. This type of core is called the \textit{strong sequential core}, and remains stable at every possible state. We investigate its characteristics, including the non-emptiness and optimality. Moreover, we prove that in the developed task allocation game, the strong sequential core is equivalent to the \textit{Walrasian equilibrium} under uncertainty; i.e., a steady state which is achieved even if the state is a priori unknown. Consequently, it can be implemented by using the \textit{Walras' tatonnoment} process, also known as the \textit{Walrasian auction}. To the best of our knowledge, the developed analytical framework appears in the literature for the first time. Our work extends the state-of-the-art in the following aspects: 
\begin{itemize}
\item Our system model is general and consistent with real-world scenarios, as explained in the following: (i) For every 
      CPS (autonomous agent), the task arrival is random. Tasks are inhomogeneous (transmission, computation, measurement, etc.) and 
			impose different demands such as requirement for resource usage (power, memory, etc.); (ii) The CPSs are distinct in types 
			and capabilities (storage capacity, CPU cycle, available radio resources, etc.); (iii) Each CPS has its own preferences in 
			selecting jobs, determined by utility scores; (iv) Tasks are dividable and can be performed by multiple CPSs each; (v) Network's 
			state is a random variable and a priori unknown. In conclusion, the model accommodates most features of complex systems, also 
			those involving human agents. Table \ref{Tb:Comparison} compares the features of our system model with state-of-the-art literature. 
\item The allocation method can be implemented in a distributed manner, which is important in particular in conjunction 
      with the generality of our system model, in particular, the involved uncertainty. As it is shown by some previous works, a 
			distributed implementation imposes less complexity and overhead compared to fully-centralized schemes. 
\item The allocation is efficient at both individual and network levels, despite limited information availability and 
      the presence of randomness/uncertainty. In addition, the solution is stable, in the sense that no individual benefits from 
			not accepting the proposal to join the grand coalition initially or by a unilateral deviation from it after the allocation. 
			Thus all CPSs cooperate in sharing the cost and executing the jobs arrived at the network. Therefore, the approach is applicable 
			also when CPSs belong to independent stakeholders.  
\item The proposed method results in self-enforcing allocation. This means that no CPS benefits by deviating from 
      the agreements, even after the resolution of the uncertainty. Therefore, they are binding also in the absence of a central 
			authority that obligates the CPSs to follow the agreements. 
\item The developed analytical framework can be used in conjunction with a large class of utility functions that satisfy some 
      fairly common conditions. Thus the model and solution can be applied to a wide range of resource allocation problems beyond 
			task management.  
\end{itemize}
\begin{table*}[ht]
\centering
\footnotesize
\caption{Comparison of State-of-the-Art System Models}
\begin{tabular}{ |c|c|c|c|c|c|}
\hline
\backslashbox{Reference}{Feature} & Systems' Types & Task's Type & Random Network's State & Random Task Arrival & Distributed \\ \hline 
\cite{Wang14:CTA}            & Arbitrary  & Arbitrary & \xmark & \xmark & \cmark  \\ \hline
\cite{deWeerdt12:MTA}        & Arbitrary  & Arbitrary & \xmark & \xmark & \cmark  \\ \hline
\cite{Zhao16:AHD}            & Arbitrary  & Arbitrary & \xmark & \xmark & \xmark  \\ \hline
\cite{Ye16:PBSW}             & Arbitrary  & Identical & \xmark & \xmark & \cmark  \\ \hline
\cite{He17:LAT}              & Arbitrary  & Identical & \xmark & \xmark & \xmark  \\ \hline
\cite{Zhao09:EGT}            & Arbitrary  & Identical & \xmark & \xmark & \cmark  \\ \hline
\cite{Manisterski06:FEA}     & Arbitrary  & Arbitrary & \xmark & \xmark & \xmark  \\ \hline
\cite{Abdallah06:LTA}        & Arbitrary  & Arbitrary & \xmark & \xmark & \cmark  \\ \hline
\cite{Grosu02:LBD}           & Arbitrary  & Arbitrary & \xmark & \xmark & \xmark  \\ \hline
\cite{Birje12:RPS}           & Identical  & Identical & \xmark & \xmark & \xmark  \\ \hline
Our Work                     & Arbitrary  & Arbitrary & \cmark & \cmark & \cmark  \\ \hline
\end{tabular}
\label{Tb:Comparison}
\end{table*}
\subsection{Organization}
\label{subsec:Organization}
The rest of the paper is organized as follows. Section \ref{sec:SysModel} presents the system model and basic assumptions. In Section 
\ref{subsec:Examples}, we describe an exemplary problem to clarify the application of the considered system model. In Section 
\ref{sec:Start}, we formulate the task allocation problem in the presence of random task arrival and state uncertainty. Moreover, we provide a general overview of the proposed solution scheme. We introduce multi-state stochastic cooperative games in Section \ref{sec:StartT}. In addition, we describe the concept of certainty equivalence. We model the task allocation problem as a two-stage stochastic game with a priori unknown states and derive its deterministic equivalent. The state uncertainty is addressed in Section \ref{sec:Walras}, where we characterize the strong sequential core of the formulated task allocation game. In Section \ref{sec:WalrasTan}, we show that the strong sequential core can be implemented using the Walras' tatonnement process. Section \ref{sec:NumG} includes numerical results. Section 
\ref{sec:Conc} summarizes the paper and adds some concluding remarks. 
\section{System Model}
\label{sec:SysModel}
Throughout the paper, by \textit{cyber-physical system} (CPS), we refer to an autonomous entity consisting of physical and computational elements, which is capable of performing a variety of tasks, possibly at different efficiency levels. This includes measurement, computation, signal processing, and transmission.

Consider a network of $N$ CPSs, gathered in a set $\mathcal{N}$. The network can be in one of the $S$ different states, modeled as the outcomes of some random variable with some arbitrary distribution. The state is unknown a priori.\footnote{The model can be explained by considering the network as a player whose strategy set is the set of states. At every round, the player selects one of the available actions randomly, rather than strategically, based on the payoff. In the language of game theory, such player is referred to as the \textit{nature.}}~The set of network's states, $\mathcal{S}$, is finite, mutually exclusive, exhaustive, and known to all CPSs. The CPSs engage in executing some divisible tasks of $M$ different types, collected in a set $\mathcal{M}$. For $S$ states of the network and $M$ tasks, there exist $MS$ \textit{state contingent} tasks. In simple words, state contingency means that a task $m$ in state $s \in \mathcal{S}$ is regarded as another task, say $m'$, when the state changes to $s' \in \mathcal{S}$. It is worth mentioning that the analysis remains valid also if the number of task types is state-dependent; That is, if at every state $s \in \mathcal{S}$, the CPSs engage in performing $M^{(s)}$ different tasks. In such case there exists $\sum_{s \in \mathcal{S}}M^{(s)}$ state contingent tasks.

At every state $s \in \mathcal{S}$, each CPS $n \in \mathcal{N}$ is characterized by a \textit{performance index} (or \textit{type}) vector 
$\boldsymbol{\rho}_{n}^{(s)}=\left(\rho_{n1}^{(s)},...,\rho_{nM}^{(s)} \right)$. Each element $\rho_{nm}^{(s)} \in \mathcal{R}_{>0}-\{\infty\}$ of the type vector represents the ability of CPS $n \in \mathcal{N}$ to perform a task of type $m \in \mathcal{M}$ at state 
$s \in \mathcal{S}$, in the sense that larger performance index implies higher efficiency. Naturally, $\rho_{nm}^{(s)}$ depends on a variety of factors such as the quality of the available transmission channel, computational capacity, measurement precision, and the like. Note that every CPS knows its own type at every possible state, but does not know which state will be realized in future. In the following we provide an intuitive example to clarify the model.

\begin{figure}[ht]
\centering
\includegraphics[width=0.40\textwidth]{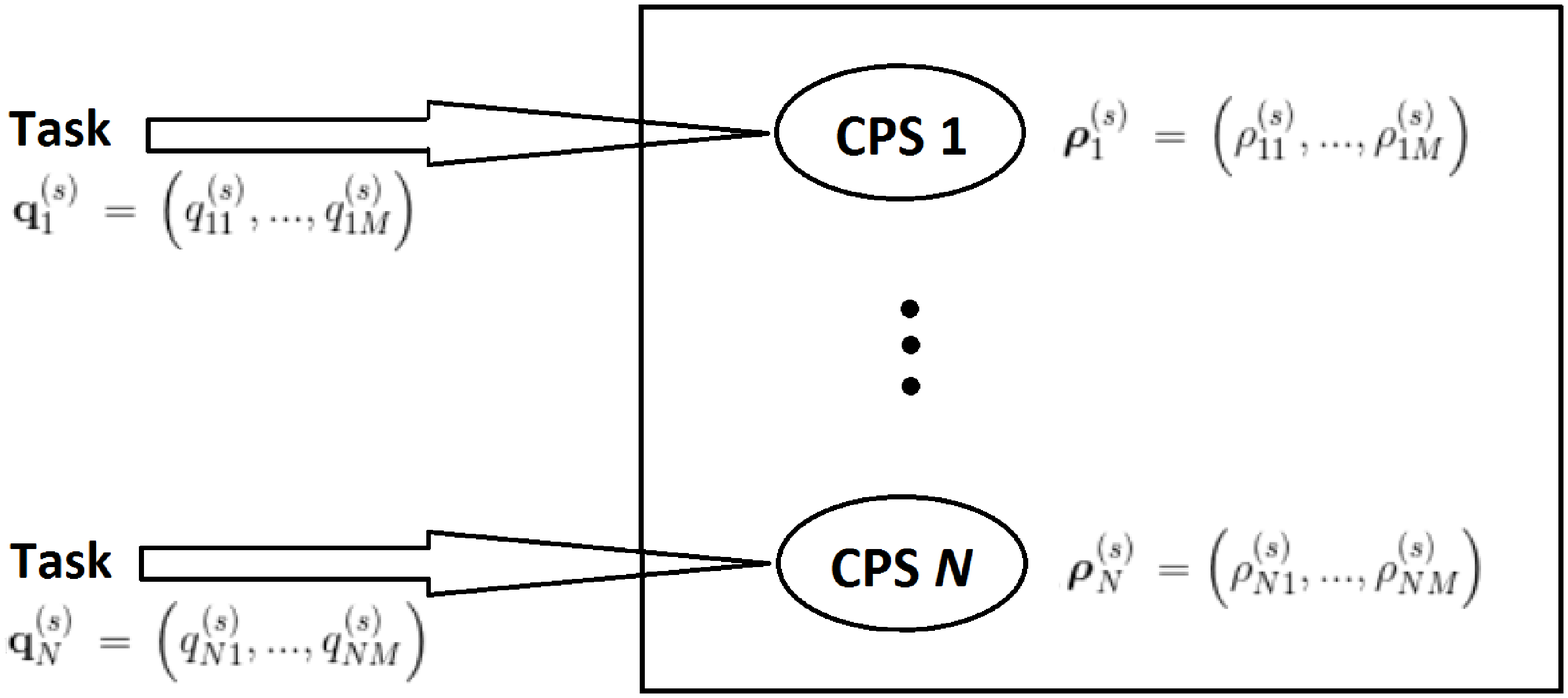}
\caption{Network of cyber-physical systems at some state $s \in \mathcal{S}$. At every state $s$, each CPS $n \in \mathcal{N}$ is characterized by a \textit{type} (efficiency) vector $\boldsymbol{\rho}_{n}^{(s)}=\left(\rho_{n1}^{(s)},...,\rho_{nM}^{(s)}\right)$, and receives the initial task vector $\mathbf{q}_{n}^{(s)}=\left(q_{n1}^{(s)},...,q_{nM}^{(s)}\right)$. The total load of every task in the network follows from (\ref{eq:Sum}). The tasks are then redistributed as described throughout the paper and summarized in 
\textbf{Fig. \ref{Fig:Chart}}.}
\label{Fig:Model}
\end{figure}

\begin{figure}[ht]
\centering
\includegraphics[width=0.30\textwidth]{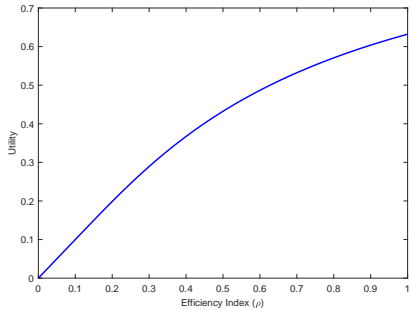}
\caption{Relation between the utility ($u(\cdot)$) and one-dimensional efficiency index (type, $\rho$), for a fixed amount of task $x$. It can be seen that larger efficiency index results in larger utility. This implies that the CPSs with the utility function 
(\ref{eq:MachUtil}) attempt to receive a higher share of the tasks in which they are more efficient, thereby contributing to the overall network efficiency.}
\label{Fig:Eff}
\end{figure}

At state $s \in \mathcal{S}$, every CPS $n \in \mathcal{N}$ randomly receives $q_{nm}^{(s)}$ unit(s) of task $m \in \mathcal{M}$ to perform. We have $\mathbf{q}_{n}^{(s)}=\left(q_{n1}^{(s)},...,q_{nM}^{(s)}\right)$, and for all $m \in \mathcal{M}$,
\begin{equation}
\label{eq:Sum}
Q_{m}^{(s)}=\sum_{n\in \mathcal{N}}q_{nm}^{(s)}.
\end{equation}
That is, $Q_{m}^{(s)}$ is the total load of task $m \in \mathcal{M}$ at state $s \in \mathcal{S}$ in the network. At every state $s \in \mathcal{S}$, the systems redistribute $Q_{m}^{(s)}$, $m \in \mathcal{M}$, among themselves to perform.
\begin{assumption}
\label{ass:InitV}
We assume that each CPS has a positive initial load of each task; that is, $q_{nm}^{(s)}>0$ for all $n \in \mathcal{N}$, $m \in 
\mathcal{M}$, and $s \in \mathcal{S}$.
\end{assumption}
For each CPS $n \in \mathcal{N}$ and at every state $s\in \mathcal{S}$, the initial arrived load of type $m \in \mathcal{M}$ is a random variable that follows an exponential distribution with parameter $\lambda_{nm}^{(s)}$. The CPSs are connected to each other (for example via internet), and the cost of exchanging data between the systems is negligible. Consequently, CPSs are able to cooperate in performing all tasks by redistributing $Q_{m}^{(s)}$, $m \in \mathcal{M}$. The system model is depicted in \textbf{Fig. \ref{Fig:Model}}. 
\begin{definition}[Allocation]
\label{de:ConCom}
An allocation of state contingent task $ms$ corresponds to an exhaustive (i.e., without any residue) division of 
$Q_{m}^{(s)}$ between $N$ CPSs. We denote any allocation by $\mathbf{x}_{n}^{(s)}=\left(x_{n1}^{(s)},...,x_{nM}^{(s)}\right)$, where 
$x_{nm}^{(s)} \in \mathcal{R}_{\geq 0}$ describes the share of task $m \in \mathcal{M}$ that should be performed by CPS $n \in 
\mathcal{N}$, if and only if state $s \in \mathcal{S}$ occurs.
\end{definition}
Note that we investigate one round of task allocation; As a result, throughout the paper, we do not use any notion of time. Naturally, the allocation profile can be used in multiple rounds of task arrival. Upon performing a share $x_{nm}^{(s)}$ of any task $m \in \mathcal{M}$ at state $s \in \mathcal{S}$, a CPS $n \in \mathcal{N}$ receives a utility denoted by $u_{nm}^{(s)}\left(x_{nm}^{(s)}\right)$. Note that as the task arrival is random, $x_{nm}^{(s)}$ is random as well. We model the network by a multi-agent system, where each CPS is a risk-averse player. The assumption of risk aversion, modeled by a concave utility function, is widely used in multi-agent systems \cite{Maghsudi17:DUAEH}, \cite{Ranadheera18:MGA}. We capture the risk-aversion by an exponential utility function as\footnote{Note that any cost-sharing game can be modeled similarly, by considering costs as negative utility. Note that other functions can be used to model the utility as long as some regularity conditions are satisfied.}
\begin{equation}
\label{eq:MachUtil}
u_{nm}^{(s)}\left(x_{nm}^{(s)}\right)=\rho_{nm}^{(s)}\left(1-e^{-\frac{1}{\rho_{nm}^{(s)}}x_{nm}^{(s)}}\right).
\end{equation}
\textbf{Fig. \ref{Fig:Eff}} shows the relation between the utility and the efficiency for a given amount of task. Intuitively, this choice of utility function guarantees that a CPS with higher efficiency makes larger utility by performing a specific share of a given task, compared to a CPS with lower efficiency. Thus, in a Pareto-optimal task allocation, CPSs with better performance index tend to receive more tasks to perform, which improves the efficiency of the platform. Note that the utility involves the cost of performing task implicitly through types or performance indexes. In other words, high performing cost adversely affects the performance index, which reduces the utility. Exponential utility function is also used by some previous works to model the satisfaction level of the agents in multi-agent systems. Examples include \cite{Khan18:WUF}, \cite{Chen09:MRA}, \cite{He10:OOI}, \cite{Shrestha08:MPU}.

The total utility follows as 
\begin{equation}
\label{eq:UtilityOne}
u_{n}^{(s)}\left(\mathbf{x}_{n}^{(s)}\right)=\sum_{m\in \mathcal{M}}u_{nm}^{(s)}\left(x_{nm}^{(s)}\right), 
\end{equation}
where $\mathbf{x}_{n}^{(s)}=\left(x_{n1}^{(s)},...,x_{nM}^{(s)}\right)$. By (\ref{eq:UtilityOne}), the utility is task-separable. 
\begin{proposition}
\label{pr:UtiAss}
For every $n \in \mathcal{N}$, the utility function defined by (\ref{eq:UtilityOne}) satisfies the following:
\begin{enumerate}[(a)]
 \item At every state $s \in \mathcal{S}$, $u_{n}^{(s)}$ is continuous, concave and monotonically increasing;
 \item The total utility function $u_{n}=\sum_{s\in\mathcal{S}}u_{n}^{(s)}$ is continuous and state-separable.
\end{enumerate} 
\end{proposition}
\begin{IEEEproof}
See Appendix \ref{SubSec:Utility}.
\end{IEEEproof}
\section{An Exemplary Application: Energy Harvesting Small Cell Network with Edge Computing}
\label{subsec:Examples}
Next generation networks are foreseen to be cognitive, meaning that the involved entities are expected to come to a consensus in distributed decision-making processes, where the outcomes affect all of them jointly. In such networks, many applications include some variables that affect (or even exclusively determine) the demands for the tasks to be done by the network and/or the performance efficiency of the involved entities. Although such variables play a major role in optimal decision-making, sometimes they are unknown at the time of negotiations. In such circumstances, we propose to consider every possible scenario with respect to the unknown variables as a state. Therefore, the state is a random variable whose realization is determined simply at random, by a (virtual) player referred to as the \textit{nature}. Moreover, there are many circumstances in which the utility of each network entity is not deterministic, but rather a random variable. The randomness in the utility function arises as a result of randomness in numerous characteristics of the network. In the following, we provide a high-level example of the proposed system model, in order to demonstrate its general applicability. In Section \ref{sec:NumToy}, we provide numerical analysis for the described scenario.

Consider a heterogeneous wireless small cell network with a set $\mathcal{N}$ of small base stations (SBS). Driven by the ever-increasing demand for delay-sensitive computation and cost-effective control, each SBS is foreseen to be an intelligent unit, able to perform low-complexity computation, transmission and decision-making at the network's edge. As mentioned in Section \ref{sec:SysModel}, here we define a CPS as an engineered entity in which computation, communication and control units are integrated. From this perspective, in this application scenario, we consider each SBS as a CPS. Since the SBSs are able to communicate with each other, we observe a set of them as a network of CPSs. Primarily, each SBS acts as an amplify-and-forward relay to improve the transmission performance. As it is conventional in small cell networks, each SBS has access to a frequency channel that can be used only if it is idle. The status of the frequency channel 
(busy, idle) is determined by the activity of the primary users, therefore it is a random variable. For every SBS $n \in \mathcal{N}$, the availability of transmission channel is modeled by a Bernoulli random variable with parameter (success probability) $\alpha_{n} \in (0,1]$. Beside transmission, the SBSs act as edge servers to perform some low-load computational tasks, if enough energy is available. Thus, in this exemplary application, two tasks are divided among SBSs, namely transmission over idle channels and computation at the edge. As such, the relevant SBS's characteristics include the availability of idle transmission channels as well as the amount of energy resource.

The SBSs are deployed randomly. The random deployment increases the coverage, but it eliminates the access to the power grid. As a result, every SBS uses the power from a built-in battery to perform transmissions, which is its main duty. Moreover, it uses energy harvesting units to possibly obtain some additional energy for computational tasks. Since energy harvesting depends on the weather that is random in nature, the amount of available energy for computation is random. We consider a set $\mathcal{S}$ of possible states for the weather. At every state, the energy arrival at every SBS approximately follows a normal distribution with mean $\mu_{n}^{(s)}>0$ \cite{Maghsudi17:DUA}. Moreover, some user devices have built-in (micro) solar cells and are capable of ambient energy harvesting in addition to accessing the battery as a source of energy.

The transmission requests are always forwarded to the SBSs, irrespective of the weather. We denote the (initial) transmission requests to every SBS $n \in \mathcal{N}$ by $q_{n1}^{(s)}$, and model it by a Poisson process with rate $\lambda_{n1}>0$ for all $s \in \mathcal{S}$. In contrast, the intensity of the offloading demands depends on the weather. In the sunny weather, the devices might use their own harvested energy for computation, thereby reducing the demand for computation offloading; However, if the weather is windy, the devices would prefer to offload the computation to the SBSs. We denote the (initial) computation offloading demands by $q_{n2}^{(s)}$, and model it by a Poisson process with rate $\lambda_{n2}^{(s)}>0$, $s \in \mathcal{S}$.  

Naturally, every SBS or server is reimbursed on the basis of the quality of service (QoS) that it provides to the users. As a result, we model the utility of each SBS with (\ref{eq:MachUtil}), while assuming that $\rho_{nm}$ is directly proportional to the resource availability, based on which the QoS is determined. Intuitively, the SBSs with a larger $\alpha_{n}$ can be more successful in addressing the transmission requests, since they are more likely to have access to an idle channel. Therefore, we model the utility of SBS 
$n \in \mathcal{N}$ with the utility function given in (\ref{eq:MachUtil}), assuming that $\rho_{n1}^{(s)}=\alpha_{n}$ for all $s \in \mathcal{S}$. Furthermore, it is natural that the utility of computation in every state depends on the energy-availability in that state. Consequently, we model the utility of SBS $n \in \mathcal{N}$ with the utility function given in (\ref{eq:MachUtil}), assuming that 
$\rho_{n2}^{(s)}=\mu_{n}^{(s)}$, for every $s \in \mathcal{S}$.  
\section{The Task Allocation Problem}
\label{sec:Start}
Under state uncertainty, the distributed task management can be described in two consecutive stages: (i) At Stage $0$, the state of the network is not known. However, all agreements among CPSs with regards to task distribution are made at this stage; (ii) At Stage $1$, the state is revealed and the agreements related to the realized state becomes valid. All other agreements are void. The sequence of events is summarized in \textbf{Algorithm \ref{alg:ArDeSe}}. This model corresponds to the two-stage economy model \cite{Debreu87:TTV}. 
\begin{algorithm}
\caption{Sequence of Events under State Uncertainty}
\label{alg:ArDeSe}
\small
\begin{algorithmic}[1]
\STATE At Stage $0$, the network's state is unknown. Commitments are however made in this stage, i.e., under state uncertainty. This 
       means that, at Stage $0$, CPSs make commitments (to form coalitions, perform a share of a task, etc.) for every possible state.
\STATE At Stage $1$, the state $s \in \mathcal{S}$ is revealed. The agreements for the realized state $s$ are executed, and all 
       others become void.
\end{algorithmic}
\end{algorithm}

Let $\mathbf{a}_{n}=\left(a^{(1)}_{n},...,a^{(S)}_{n}\right)$ indicate a probability distribution over the set of states $\mathcal{S}$, where $a^{(s)}_{n}$, $s \in \mathcal{S}$, is the likelihood of occurrence of state $s \in \mathcal{S}$, as predicted by CPS $n \in 
\mathcal{N}$. Naturally, $\sum_{s \in \mathcal{S}}a^{(s)}_{n}=1$. The expected utility then yields
\begin{equation}
\label{eq:ExpUtiliyOne}
v_{n}\left(\left[\mathbf{x}_{n}^{(s)}\right]_{s\in \mathcal{S}}\right)=\sum_{s \in \mathcal{S}}a^{(s)}_{n}u_{n}^{(s)}
\left(\mathbf{x}_{n}^{(s)}\right),
\end{equation}
where $\left[\mathbf{x}_{n}^{(s)}\right]_{s\in \mathcal{S}}$ is the collection of $\mathbf{x}_{n}^{(s)}$ for all $s \in \mathcal{S}$. Note that $\mathbf{a}_{n}$ is not necessarily known by CPSs. In case $\mathbf{a}_{n}$ is unknown, some probability distribution, for instance the uniform distribution, can be used. Moreover, the CPSs do not need to agree on a specific distribution $\mathbf{a}$.\footnote{The effect of private prior on equilibrium is discussed in \cite{Colell85:MT}.}~In (\ref{eq:ExpUtiliyOne}), two random variables can be observed: tasks' loads and network's state.

In order to formalize the cooperative task allocation, we model the CPS network with an stochastic exchange economy, also under state uncertainty. The model consists in a set of consumers $n \in \mathcal{N}$ (representing CPSs), a set of divisible commodities $m \in 
\mathcal{M}$ (representing tasks), and a set of states $\mathcal{S}$. The exchange economy model captures the idea of exchanging goods, without production, where the allocation of a given amount of each commodity implies its final consumption, associated with some utility score. Since utilities and initial endowments are the main blocks of an exchange economy, we denote the exchange economy model of task allocation as $\Omega:\left\{\left[u_{n}^{(s)}\right]_{s \in \mathcal{S}},\left[\mathbf{q}_{n}^{(s)}\right]_{s \in \mathcal{S}}\right\}_{n \in \mathcal{N}}$. In a conventional model of exchange economy, all variables are deterministic and all information is provided to all agents a priori; in contrast, in our model, the nature (network) can have different states, modeled as the outcomes of some random variable, unknown a priori. Moreover, parameters such as initial endowments are random variables and their statistical characteristics very over states.  

\begin{figure}[ht]
\centering
\includegraphics[width=0.21\textwidth]{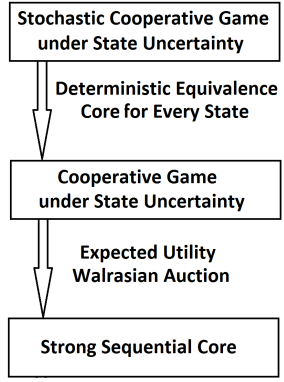}
\caption{A diagram of the proposed solution for the task allocation problem. Briefly, we first replace the stochastic state game with its deterministic equivalent. Afterward, we use the expected utility and Walrasian auction model to implement the strong sequential core.}
\label{Fig:Chart}
\end{figure}

In order to describe the redistribution (exchange) mechanism, we use \textit{virtual} prices to quantify the value of each task based on its popularity over CPSs. In this virtual model, a price $p_{m}^{(s)}$ has to be paid for every unit of a state contingent task $ms$, $m \in \mathcal{M}$ and $s \in \mathcal{S}$. We denote the price vector by $\mathbf{p}=\left[\mathbf{p}^{(s)}\right]_{s \in \mathcal{S}}$, where 
$\mathbf{p}^{(s)}=\left(p_{1}^{(s)},...,p_{M}^{(s)}\right) \in \mathcal{R}_{>0}^{M}$. Given prices, for every $n \in \mathcal{N}$, the initial endowments maps to a budget set as
\begin{equation}
\label{eq:BudSet}
\mathcal{B}_{n}^{(s)}\left(\mathbf{p}^{(s)}\right)=\left \{\mathbf{x}_{n}^{(s)}:\mathbf{p}^{(s)}\cdot \mathbf{x}_{n}^{(s)} \leq 
\mathbf{p}^{(s)} \cdot \mathbf{q}_{n}^{(s)} \right\}.
\end{equation}
The CPSs are expected-utility maximizer; that is, at stage 0, every CPS $n \in \mathcal{N}$ would like to agree on its state-dependent load of tasks $\left[\mathbf{x}_{n}^{(s)}\right]_{s\in \mathcal{S}}$ so as to maximize its expected utility. Let $\mathcal{X}_{n}^{(s)}$ be the set of all possible demands for CPS $n \in \mathcal{N}$. Each CPS $n \in \mathcal{N}$ solves the following optimization problem: 
\begin{equation}
\label{eq:UserOpt}
\begin{aligned}
&\underset{\left[\mathbf{x}_{n}^{(s)}\right]_{s \in \mathcal{S}} \in \mathcal{X}_{n}^{(s)}}{\textup{maximize}} ~~v_{n}
\left(\left[\mathbf{x}_{n}^{(s)}\right]_{s \in \mathcal{S}}\right)\\ 
&\textup{s.t.}~~\mathbf{p}^{(s)} \cdot \mathbf{x}_{n}^{(s)}\leq \mathbf{p}^{(s)} \cdot \mathbf{q}_{n}^{(s)},~s \in \mathcal{S}.
\end{aligned}
\end{equation}
Let $\mathcal{X}$ be the set of all possible task allocations. From the platform's perspective, the objective is to maximize the aggregate utility performance (social welfare), i.e.,
\begin{equation}
\label{eq:ExpUtiliy}
\begin{aligned}
&\underset{\left[\mathbf{x}_{n}^{(s)}\right]_{s\in \mathcal{S}, n \in \mathcal{N}} \in \mathcal{X}}{\textup{maximize}}~~
\sum_{n\in\mathcal{N}}v_{n}\left(\left[\mathbf{x}_{n}^{(s)}\right]_{s\in \mathcal{S}}\right)\\
&\textup{s.t.}~~\sum_{n \in \mathcal{N}} \mathbf{x}_{n}^{(s)}\leq \sum_{n \in \mathcal{N}}\mathbf{q}_{n}^{(s)},~s \in \mathcal{S}.
\end{aligned}
\end{equation}
In the rest of this paper, we model and solve the task allocation problem using stochastic cooperative games under state uncertainty. The procedure is summarized in \textbf{Fig. \ref{Fig:Chart}}. Briefly, we first replace the stochastic state game with its deterministic equivalent. Afterward, we use the expected utility and Walrasian auction model to implement the strong sequential core. Details are discussed in the upcoming sections.
\section{Stochastic Cooperative Games in Two-Stage Economies}
\label{sec:StartT}
As mentioned before, in our model, CPSs are connected in a network. Consequently, they are able to share data and thereby to cooperate in performing tasks. Hence we solve the task management problem using a stochastic cooperative game model under state uncertainty, where players form coalitions, engage in performing joint actions, and each receive some utility, all given no prior information about the future state and/or arrived jobs. Let $\mathcal{C}$ indicate the set of all possible coalitions among $N$ CPSs with its cardinality being $2^{N}$. The state game, i.e., the game in any state $s \in \mathcal{S}$, is defined as a tuple $\mathfrak{G}^{(s)}:\left(\mathcal{N},\left[\mathbf{w}_{c}^{(s)}\right]_{c \in \mathcal{C}},\left[u_{n}^{(s)}\right]_{n\in \mathcal{N}}, \left[\geqslant_{n}\right]_{n \in \mathcal{N}}\right)$, where
\begin{itemize}
\item $\mathcal{N}$ is the set of $N$ players, here CPSs;
\item $\mathbf{w}_{c}^{(s)}$ is the characteristics function $2^{N} \to \mathcal{R}_{\geq 0}^{M}$, which assigns a value to each of the 
      $2^{N}$ coalitions $c \in \mathcal{C}$ at state $s \in \mathcal{S}$. This implies that the value of each coalition is state-dependent;
\item $u_{nm}^{(s)}$ is the utility function of player $n \in \mathcal{N}$ by performing task $m \in \mathcal{M}$ at state $s \in 
      \mathcal{S}$;			
\item $\geqslant_{n}$ is the preference relations of player (CPS) $n \in \mathcal{N}$ over the set $\mathcal{R}_{\geq 0}$ of rewards.  
\end{itemize}
We define preference relation $\geqslant_{n}$ in terms of \textit{Von Neumann-Morgenstern} preference based on the utility function 
$u_{nm}(x):\mathcal{R}_{>0} \to \mathcal{R}$, where for $x,y \in \mathcal{R}_{>0}$,
\begin{equation}
\label{eq:preference}
x \geqslant_{n}y~\text{iff}~\mathbb{E}\left[u_{nm}(x)\right]>\mathbb{E}\left[u_{nm}(y)\right],
\end{equation}
where expectation is taken with respect to the random reward; That is, a stochastic reward is preferred to the other if it yields a larger utility in expectation. This definition of preference corresponds to \textit{rationality}. 

In our problem, the initial wealth of each singleton coalition consisting of CPS $n \in \mathcal{N}$ yields $\mathbf{q}_{n}^{(s)}$, so that the initial wealth of every coalition $c \in \mathcal{C}$ is given as $\sum_{n\in c} \mathbf{q}_{n}^{(s)}$. After the allocation, i.e., after determining $\mathbf{x}_{n}^{(s)}$, the value of coalition $c \in \mathcal{C}$ follows as $\mathbf{w}_{c}^{(s)}=\sum_{n \in c}
\mathbf{x}_{n}^{(s)}$. Therefore $\mathbf{w}_{c}^{(s)}$ is stochastic with finite first moment. 

Let coalition $c$ include $N_{c}$ CPSs, collected in a set $\mathcal{N}_{c}$. At state $s$, each member $n \in \mathcal{N}_{c}$ receives a task vector \cite{Suijs99:SCG} 
\begin{equation}
\label{eq:CoShare}
\mathbf{x}_{n,c}^{(s)}=\mathbf{r}_{n,c}^{(s)}\mathbf{w}_{c}^{(s)}, 
\end{equation}
where  
\begin{equation}
\label{eq:Scond}
\sum_{n \in \mathcal{N}_{c}}\mathbf{r}_{n,c}^{(s)}=1,
\end{equation}
with $\mathbf{r}_{n,c}^{(s)}\geq 0$.
The pay off of member $n \in \mathcal{N}_{c}$ then yields
\begin{equation}
\label{eq:CoShareUtility}
\mathbf{u}_{n,c}^{(s)}\left(\mathbf{x}_{n,c}^{(s)}\right)=\mathbf{f}_{n,c}^{(s)}+ \mathbf{u}_{n,c}^{(s)}\left(\mathbf{r}_{n,c}^{(s)}
\mathbf{w}_{c}^{(s)}\right), 
\end{equation}
where, for utility function (\ref{eq:MachUtil}), at any Pareto-optimal allocation it holds \cite{Suijs99:SCG}, \cite{Wilson68:ToS} 
\begin{equation}
\label{eq:Fcond}
\sum_{n \in \mathcal{N}_{c}}\mathbf{f}_{n,c}^{(s)}=0.
\end{equation}
The term $\mathbf{f}_{n,c}^{(s)}$ is the deterministic part of the payoff. It is in general used to give some coalition members higher priority, or in other words, larger rewards. In this paper we assume $\mathbf{f}_{n}^{(s)}=0$ for all $n \in \mathcal{N}$ and $s \in 
\mathcal{S}$. The term $\mathbf{r}_{n,c}^{(s)}\mathbf{w}_{c}^{(s)}$ is stochastic. For coalition $c \in \mathcal{C}$, we show the allocation profile as $\mathbf{x}_{c}^{(s)}=\left[\mathbf{x}_{n,c}^{(s)} \right]_{n\in \mathcal{N}_{c}}$. At state $s \in \mathcal{S}$, the set of all possible allocations for coalition $c \in \mathcal{C}$ is denoted by $\mathcal{X}_{c}^{(s)}$. At every state $s \in \mathcal{S}$, each CPS 
$n \in \mathcal{N}$ evaluates its payoff with some utility function, as discussed in Section \ref{sec:SysModel}.
\subsection{Certainty (Deterministic) Equivalence}
\label{sec:DetEqui}
As we mentioned in Section \ref{sec:SysModel}, the optimization problem (\ref{eq:ExpUtiliyOne}) includes two random variables: the tasks' loads and the network's states. In order to approach the problem, we first eliminate the randomness in $\mathbf{x}_{n}^{(s)}$ (tasks' loads), $n \in \mathcal{N}$. To this end, at every state, the stochastic cooperative game is transformed to a deterministic cooperative game using the concept of \textit{certainty (deterministic) equivalence} \cite{Suijs99:SCG}. In order to transform a stochastic game to a deterministic one, for each agent $n \in \mathcal{N}$, we need to specify the deterministic share of the coalition value, in a way that every player is indifferent between receiving the stochastic reward $\mathbf{x}_{n,c}^{(s)}$ and its deterministic equivalent shown by $\mathbf{d}_{n}^{(s)}\left(\mathbf{x}_{n,c}^{(s)}\right)=\left(d_{n1}^{(s)},...,d_{nM}^{(s)}\right)$. The following definition characterizes the deterministic equivalent of a stochastic game $\mathfrak{G}^{(s)}$ to guarantee such indifference as well as Pareto-optimality.\footnote{Note that the deterministic equivalent is developed independently for every task $m \in \mathcal{M}$ at every state $s \in \mathcal{S}$.}
\begin{definition}[Deterministic Equivalence \cite{Suijs99:SCG}]
\label{de:DetEqui}
Consider the stochastic game $\mathfrak{G}^{(s)}$ defined by the tuple $\left(\mathcal{N},\left[\mathbf{w}^{(s)}_{c}\right]_{c\in \mathcal{C}}, \left[u_{n}^{(s)}\right]_{n\in \mathcal{N}}\left[\geqslant_{n}\right]_{n \in \mathcal{N}}\right)$. Then the associated cooperative game with deterministic payoff is given by $\mathfrak{G}^{(s)}_{D}:\left(\mathcal{N},\left[\mathbf{w}^{(s)}_{c,D}\right]_{c\in \mathcal{C}},
\left[u_{n,D}^{(s)}\right]_{n\in \mathcal{N}},\left[\geqslant_{n}\right]_{n \in \mathcal{N}}\right)$, where 
\begin{equation}
\label{eq:DetEqTheorem}
\mathbf{w}^{(s)}_{c,D}=\underset{\mathbf{x}_{n,c}^{(s)} \in \mathcal{X}_{c}^{(s)}}{\textup{maximum}}\sum_{n\in\mathcal{N}_{c}}\mathbf{d}_{n}
\left(\mathbf{x}_{n,c}^{(s)}\right),
\end{equation}
and
\begin{equation}
\label{eq:DetEqTheoremTwo}
u_{n,D}^{(s)}=\sum_{m \in \mathcal{M}}u_{nm}^{(s)}\left(d_{nm}^{(s)}\right).
\end{equation}
Moreover, for every $n \in \mathcal{N}$, $\mathbf{d}_{n}^{(s)}\left(\mathbf{x}_{n,c}^{(s)}\right)$ in (\ref{eq:DetEqTheorem}) satisfies the following conditions:
\begin{enumerate}[(a)]
 \item $\forall~\mathbf{x}_{n,c}^{(s)} \in \mathcal{R}_{>0}^{M}: \mathbf{x}_{n,c}^{(s)} \approx_{n} 
       \mathbf{d}_{n}^{(s)}\left(\mathbf{x}_{n,c}^{(s)}\right)$;
 \item $\forall~\mathbf{x}_{n,c}^{(s)}, \mathbf{y} \in \mathcal{R}_{>0}^{M}: \mathbf{x}_{n,c}^{(s)} \geqslant_{n} \mathbf{y}$ $\iff$ 
       $\mathbf{d}_{n}^{(s)}\left(\mathbf{x}_{n,c}^{(s)}\right) \geq \mathbf{d}_{n}^{(s)}\left(\mathbf{y}\right)$;
 \item $\forall~\mathbf{k} \in \mathcal{R}^{M}$ with $\mathbf{k}$ being deterministic: $\mathbf{d}_{n}^{(s)}\left(\mathbf{k}\right)=
       \mathbf{k}$;
 \item $\forall~\mathbf{x}_{n,c}^{(s)} \in \mathcal{R}_{>0}^{M}:\mathbf{d}_{n}^{(s)}\left(\mathbf{x}_{n,c}^{(s)}-\mathbf{d}_{n}^{(s)}
       \left(\mathbf{x}_{n,c}^{(s)} \right)\right)=0$;
 \item $\forall~\mathbf{x}_{n,c}^{(s)} \in \mathcal{R}_{>0}^{M}$ and deterministic $\mathbf{k},\mathbf{k}' \in \mathcal{R}^{M}$, with 
       $\mathbf{k}<\mathbf{k}':\mathbf{d}_{n}^{(s)}\left(\mathbf{x}_{n,c}^{(s)}+\mathbf{k}\right)<\mathbf{d}_{n}^{(s)}
			 \left(\mathbf{x}_{n,c}^{(s)}+\mathbf{k}'\right)$.
\end{enumerate} 
\end{definition}
As mentioned before, $\sum_{n \in \mathcal{N}_{c}}\mathbf{f}_{n,c}^{(s)}=0$; Thus, in (\ref{eq:DetEqTheorem}), the maximum is actually taken over the set $\{\mathbf{r} \in \left[0,1 \right]^{M}| \sum_{n \in \mathcal{N}_{c}}\mathbf{r}_{n}=\mathbf{1}\}$. 
\begin{proposition}
\label{pr:exponential}
Let the utility function of a strictly risk-averse player $n \in \mathcal{N}$ be given by (\ref{eq:MachUtil}). Then the certainty equivalent of the random reward $x_{nm,c}^{(s)}$ is given by
\begin{equation}
\label{eq:utiexpDet}
\begin{aligned}
d_{nm}^{(s)}\left(x_{nm,c}^{(s)}\right)&=u_{nm,c}^{-1}\left(\mathbb{E}\left[u_{nm,c}^{(s)}\left(x_{nm,c}^{(s)}\right)\right]
\right)\\
&=-\rho_{nm}^{(s)} \ln \left(\mathbb{E} \left[e^{\frac{-1}{\rho_{nm}^{(s)}}\left(x_{nm,c}^{(s)}\right)}\right] \right).
\end{aligned}
\end{equation}
\end{proposition} 
\begin{IEEEproof}
See Appendix \ref{SubSec:exponential}.
\end{IEEEproof}
In the rest of the paper, we will analyze the stochastic two-stage game based on the notion of deterministic equivalence. 
\section{Analysis of the Two Stage Game}
\label{Sec:CoreNalysis}
In order to analyze the multi-state two-stage game, we start from its basis, which is the stochastic game at every state. For a single-state cooperative game with deterministic rewards, a well-known solution concept is the \textit{core} \cite{Gillies59:SG}. We state the formal definition of the core below.
\begin{definition}[Core]
\label{de:Core}
Consider a (deterministic) cooperative game $\mathfrak{G}:\left(\mathcal{N},\left[\mathbf{w}_{c}\right]_{c \in \mathcal{C}},[u_{n}]_{n \in \mathcal{N}},\left[\geq_{n}\right]_{n \in \mathcal{N}}\right)$. The core, denoted by $\mathbb{C}$, is a set of feasible payoff allocations for the grand coalition $\left[\bar{\mathbf{x}}_{n}\right]_{n \in \mathcal{N}}$ satisfying the following conditions:
\begin{enumerate}[(a)]
 \item Efficiency: $\sum_{n \in \mathcal{N}}\bar{\mathbf{x}}_{n,c}=\mathbf{w}_{c}$; 
 \item Coalitional rationality: There is no coalition $c \subset \mathcal{N}$ and allocation $\left[\mathbf{y}_{n}\right]_{n \in 
       \mathcal{N}}$ which is preferred by every $n \in c$ over the grand coalition and allocation $\left[\bar{\mathbf{x}}_{n}\right]_{n 
			 \in \mathcal{N}}$; formally, there is no $\left[\mathbf{y}_{n}\right]_{n \in \mathcal{N}}$ so that $u_{n}(\bar{x}_{n})\leq 
			 u_{n}(y_{n})~\forall~n \in c$ and $\exists~n \in c$ such that $u_{n}(\bar{x}_{n})<u_{n}(y_{n})$.
\end{enumerate} 
\end{definition}
In words, in a core solution of a deterministic cooperative game, all the wealth is allocated in a way that there is no coalition of players in which all members benefit by deviating from the grand coalition. As discussed before, in our setting, the state games are stochastic. In order to find the core of the stochastic state game, we use the following theorem. The theorem states the relation between the core of a stochastic game and that of its deterministic equivalent.
\begin{theorem}[\hspace{1sp}\cite{Suijs99:SCG}]
\label{th:DetEquCore}
The core of stochastic game $\mathfrak{G}^{(s)}$ is identical to the core of its deterministic equivalent $\mathfrak{G}^{(s)}_{D}$.
\end{theorem}
By Theorem \ref{th:DetEquCore}, the problem of finding the core of the stochastic game at every state boils down to finding the core of its certainty equivalent. Consequently, when analyzing the two-stage game, we can replace the stochastic state game with its deterministic equivalent. 

As mentioned before, at every state $s \in \mathcal{S}$, the random initial value of each singleton coalition including CPS $n \in 
\mathcal{N}$ equals $\mathbf{q}_{n}^{(s)}$. Moreover, the value of every coalition $c$ with $N_{c}$ members is given as $\sum_{n \in 
\mathcal{N}_{c}}\mathbf{q}_{n}^{(s)}$, so that the random value of the grand coalition yields $\mathbf{Q}^{(s)}$. Thus problem 
(\ref{eq:UserOpt}) can be transformed as 
\begin{equation}
\label{eq:DetEqivOne}
\begin{aligned}
&\underset{\left[\mathbf{r}_{n}^{(s)}\right]_{s \in \mathcal{S}}\in [0,1]^{M}}{\textup{maximize}}~~v_{n,D}\left(\left[\mathbf{r}_{n}^{(s)}\mathbf{Q}^{(s)}\right]_{s \in \mathcal{S}}\right)\\ 
&\textup{s.t.}~~\mathbf{p}^{(s)} \cdot \mathbf{r}_{n}^{(s)}\leq \mathbf{p}^{(s)} \cdot \frac{\mathbf{q}_{n,D}^{(s)}}
{\mathbf{Q}_{D}^{(s)}},~s \in \mathcal{S},
\end{aligned}
\end{equation}
where 
\begin{equation}
\label{eq:DetEqivThree}
v_{n,D}\left(\left[\mathbf{r}_{n}^{(s)}\mathbf{Q}^{(s)}\right]_{s\in \mathcal{S}}\right)=\sum_{s \in \mathcal{S}}a^{(s)}
u_{n}^{(s)}\left(\mathbf{d}_{n}^{(s)}\left[\mathbf{r}_{n}^{(s)}\mathbf{Q}^{(s)}\right]_{s \in \mathcal{S}}\right),
\end{equation}
and 
\begin{equation}
\label{eq:DetEqivFour}
\mathbf{q}_{n,D}^{(s)}=-\boldsymbol{\rho}_{n}^{(s)}\ln\left(\mathbb{E}\left[e^{\frac{-1}{\boldsymbol{\rho}_{n}^{(s)}}\left(\mathbf{q}_{n}^{(s)}\right)}\right]\right).
\end{equation}
Moreover, problem (\ref{eq:ExpUtiliy}) is converted to
\begin{equation}
\label{eq:DetEqivTwo}
\begin{aligned}
&\underset{\left[\mathbf{r}_{n}^{(s)}\right]_{s\in \mathcal{S}, n \in \mathcal{N}} \in [0,1]^{NM}}{\textup{maximize}}~~\sum_{n\in\mathcal{N}}v_{n,D}\left(\left[\mathbf{r}_{n}^{(s)}\mathbf{Q}^{(s)}\right]_{s\in \mathcal{S}}\right)\\
&\textup{s.t.}~~\sum_{n \in \mathcal{N}} \mathbf{r}_{n}^{(s)}=\mathbf{1},~s \in \mathcal{S}.
\end{aligned}
\end{equation}
%

After reformatting the game by using the concept of certainty equivalence, we face a multi-state deterministic cooperative game under state uncertainty. For such game, the conventional \textit{core} concept (Definition \ref{de:Core}) does not suffice, and new solutions are developed in the literature. This includes \textit{two-stage core}, \textit{strong sequential core}, and \textit{weak sequential core}. For a comparative study see \cite{Herings06:TWS} and \cite{Habis11:TUG}. In this paper we focus on the strong sequential core (SSC), defined below. \textit{Note that, since we eliminated the randomness by using deterministic equivalent, in the rest of the paper, $\mathbf{x}_{n}^{(s)}$ is a deterministic variable which simply maps to $\mathbf{r}_{n}^{(s)}$, as $\mathbf{x}_{n}^{(s)}=\mathbf{r}_{n}^{(s)}\mathbf{Q}^{(s)}$.} 
\begin{definition}[Strong Sequential Core \cite{Herings06:TWS}]
\label{de:WeakCore}
The strong sequential core of the deterministic game $\mathfrak{G}$ is the set of all feasible allocations $\bar{\mathbf{x}}=
\left[\bar{\mathbf{x}}_{n,\mathcal{N}}^{(s)}\right]_{s \in \mathcal{S}, n\in \mathcal{N}}$ for the grand coalition ($\mathcal{N}$), for which the following holds:
\begin{enumerate}[(a)]
 \item $\bar{\mathbf{x}}^{(s)} \in \mathbb{C}\left(\mathfrak{G}^{(s)}\right)$ for all $s \in \mathcal{S}$;
 \item There is no coalition $c \subseteq \mathcal{N}$ and allocation $\mathbf{x}_{c}$ such that $\mathbf{w}_{c}^{(s)}=\sum_{n \in c} 
       \mathbf{q}^{(s)}_{c}$ for all $s \in \mathcal{S}$, and $v_{n,c}\left(\left[\mathbf{x}_{n,c}^{(s)}\right]_{s \in 
			 \mathcal{S}}\right)>v_{n,c}\left(\left[\bar{\mathbf{x}}_{n,\mathcal{N}}^{(s)}\right]_{s \in \mathcal{S}}\right)$ for all $n 
			 \in c$.  
\end{enumerate}
\end{definition}
In words, for an allocation to belong to the strong sequential core, the following must hold:
\begin{enumerate}[(a)]
\item The allocation should belong to the core of the state game in every state $s \in \mathcal{S}$; That is, it must be stable against all the deviations \textit{ex-post} (after revealing the uncertainty).
\item No coalition of agents (including the grand coalition) shall be able to improve upon $\bar{\mathbf{x}}$ \textit{ex-ante} 
(before revealing the uncertainty). A coalition can improve upon an allocation $\bar{\mathbf{x}}$ with allocation 
$\mathbf{x}$, if $\mathbf{x}$ is feasible and gives \textit{each} CPS of the deviating coalition a higher expected utility.\footnote{The definitions of \textit{feasible allocation} and \textit{deviation} are included in the appendix (Definition \ref{de:FeasAll} and Definition \ref{de:Dev}, respectively), in order to maintain the consistency and readability.}
\end{enumerate}
Before revealing the uncertainty in the state of the network, every agent joins/leaves coalitions with the goal of maximizing its expected utility. Nonetheless, a coalition which has no incentive to block a specific allocation ex-ante, may object against it ex-post, i.e., after the uncertainty is resolved. Mostly, it is assumed that the agents remain committed to the agreements also after the resolution of uncertainty, but clearly such assumption is restrictive and in general difficult to implement. The strong sequential core, however also guarantees ex-post commitment, which makes it desirable as a solution concept.

In the next section we describe a method to implement the strong sequential core, and we establish some of its characteristics, including the non-emptiness, in our designed task allocation game.
\section{Walrasian Equilibrium under Uncertainty and Strong Sequential Core}
\label{sec:Walras}
Till now, we showed that some stochastic cooperative games can be transformed into deterministic games using the notion of \textit{certainty equivalence} (Definition \ref{de:DetEqui}). If the game is played under uncertainty, i.e., if the nature randomly takes one of some a priori unknown states, the deterministic equivalent should be calculated separately for every possible state. The resulted game is then a multi-state deterministic cooperative game, but with state uncertainty. As discussed before, a general solution concept for such games is the \textit{strong sequential core} (Definition \ref{de:WeakCore}). However, similar to the conventional core concept developed for games without uncertainty, it is necessary to characterize such equilibrium, for instance to establish its non-emptiness in the game under investigation. Upon existence, it should be discussed that how such equilibrium can be implemented.  

For deterministic cooperative games without uncertainty, it is known that Walrasian equilibrium lies in the core of the corresponding cooperative game \cite{Gul99:WEGS}. In other words, with implementing a Walrasian equilibrium, one achieves a core solution. In what follows, we first provide the definition of Walrasian equilibrium under state uncertainty, and then describe its characteristics. Afterward, we establish that under few reasonable assumptions, Walrasian equilibrium under uncertainty belongs to the strong sequential core of the corresponding deterministic cooperative game with state uncertainty. Moreover, since we have used certainty equivalence, it is also a core solution for the initial stochastic cooperative game under uncertainty. We also describe how to implement this solution.
\subsection{Exchange Economy and Walrasian Equilibrium under Uncertainty}
\label{subsec:Walras}
In an exchange economy under uncertainty, the equilibrium notion is the \textit{Walrasian equilibrium under uncertainty}, also called \textit{Arrow-Debreu equilibrium}, defined in the following.
\begin{definition}[Walrasian Equilibrium under Uncertainty]
\label{de:ArDeEq}
A set of allocation matrices $\bar{\mathbf{x}}^{(s)}$, together with a price vector $\mathbf{p}^{(s)}$, $s \in \mathcal{S}$, are Walrasian equilibrium under uncertainty if 
\begin{enumerate}[(a)]
\item $\forall~n \in \mathcal{N}$, $\left[\bar{\mathbf{x}}_{n}^{(s)}\right]_{s \in \mathcal{S}}$ maximizes $v_{n}(\cdot)$ on 
      $\mathcal{B}_{n}$;
\item Market clears, that is, for all $s \in \mathcal{S}$
  \begin{equation}
  \label{eq:FeasThrMC}
  \sum_{n \in \mathcal{N}}\bar{\mathbf{x}}_{n}^{(s)}=\sum_{n \in \mathcal{N}}\mathbf{q}_{n}^{(s)}.
  \end{equation}  
\end{enumerate}
\end{definition}
In the following, we investigate the existence and characteristics of Walrasian equilibrium under uncertainty, in our designed task allocation economy. 
\begin{proposition}
\label{Pr:WalChar}
In the deterministic equivalent of our designed multi-state stochastic task exchange economy, Walrasian equilibrium under state uncertainty exists. Moreover, it is unique, Pareto-efficient, and thus social-optimal. 
\end{proposition}
\begin{IEEEproof}
See Appendix \ref{SubSec:PropUtiExpN}.
\end{IEEEproof}
\subsection{Relation to The Strong Sequential Core}
\label{subsec:Relation}
It is obvious that in an exchange economy with no uncertainty, the definition of Walrasian equilibrium collapses to that of Walrasian equilibrium under uncertainty when only one state exists. The following theorem declares that in such economy, the Walrasian equilibrium lies inside the core.
\begin{theorem}[\hspace{1sp}\cite{Debreu63:ALT}]
\label{th:LieCore}
Let allocation $\bar{\mathbf{x}}$ together with price vector $\mathbf{p}$ be a Walrasian equilibrium for an exchange economy 
$\Omega:\left\{u_{n},\mathbf{q}_{n}\right\}_{n \in \mathcal{N}}$ (with no state uncertainty), where $\bar{\mathbf{x}}=
\left[\bar{\mathbf{x}}_{n}\right]_{n \in \mathcal{N}}$ and $\bar{\mathbf{x}}_{n}=\left(\bar{x}_{n1},...,\bar{x}_{nm} \right)$. 
If each $\bar{x}_{nm}$, $n \in \mathcal{N}$ and $m \in \mathcal{M}$, is locally non-satiated (LNS), then $\bar{\mathbf{x}}$ lies in the core of $\Omega$.
\end{theorem}
The definition of \textit{locally non-satiated} can be found in Appendix \ref{subsec:TechPreEx}, Definition \ref{de:LoSatis}. Informally, 
non-satiation means that greater quantities provide higher levels of satisfaction to individuals. In should be mentioned that 
non-satiation is implied by the monotonicity, whereas the reverse does not hold. The following theorem states the conditions under which any Walrasian equilibrium with state uncertainty lies in the strong sequential core. 
\begin{theorem}
\label{th:WalCore}
In our designed multi-state stochastic task exchange economy, the Walrasian equilibrium $\left(\bar{\mathbf{x}}, \mathbf{p}\right)$ lies in strong sequential core.
\end{theorem}
\begin{IEEEproof}
See Appendix \ref{SubSec:ProofThOne}.
\end{IEEEproof}
\section{Implementing the Strong Sequential Core: Walras' Tatonnement Process}
\label{sec:WalrasTan}
In the previous section, we showed that Walrasian equilibrium under uncertainty lies in strong sequential core. Thus, in order to implement a core solution, it suffices to implement a Walrasian equilibrium. 

In a competitive market, every self-interested agent selects its demand so as to maximize its own utility score. Such selfish behavior yields a conflict which degrades the network's performance. Consequently, a mechanism should be used to guide the agents to a stable and efficient operating point. One such mechanism is the \textit{Walras' tatonnement} process, also called the \textit{Walrasian auction}. The process requires a coordinator (auctioneer), which, at each round, announces the prices, starting at some random initial point. Afterward, agents disclose their demands at the given prices, so that the auctioneer adjusts prices to claimed demands. The process continues until the market clears; that is, when a set of prices yields a demand equal to supply. At this point, prices and demands are final, and the auction process terminates, i.e., trade occurs \cite{Uzawa60:WTT}. Let $z \left(\mathbf{p}\right)$ be the excess demand given price vector 
$\mathbf{p}$, given by
\begin{equation}
\label{eq:Execss}
z_{m}^{(s)}(\mathbf{p})=\sum_{n\in \mathcal{N}}x^{(s)}_{nm}(\mathbf{p}, \mathbf{p}.\mathbf{q}_{n})-\sum_{n \in \mathcal{N}}q^{(s)}_{nm}.
\end{equation}
The price adjustment rule is given by \cite{Levin06:GE}
\begin{equation}
\label{eq:AuctionO}
\mathbf{p}^{(s)}(t+1)= \mathbf{p}^{(s)}(t)+\alpha z \left(\mathbf{p}^{(s)}(t)\right),
\end{equation}
for a sufficiently small $\alpha>0$. Note that the variable $t$ is a local variable that counts the number of iterations of the auction process; i.e., the number of negotiation rounds until the systems achieve some agreement on the task allocation. Clearly, the only stationary points of this process are prices $\mathbf{p}$ at which $z(\mathbf{p})=0$, i.e., equilibrium prices \cite{Levin06:GE}. The procedure is summarized in \textbf{Algorithm \ref{Alg:Auction}}. 
\begin{algorithm}
\caption{Walrasian Auction}
\label{Alg:Auction}
\small
\begin{algorithmic}[1]
\STATE Select the price adjustment factor $\alpha \to 0$.
\STATE Initialize the price of each commodity, $p_{m}^{(s)} \to 0$, $m \in \mathcal{M}$ and $s \in \mathcal{S}$.
\REPEAT
\STATE {
\begin{itemize}
 \item The auctioneer announces the prices.
 \item Each CPS declares its demand.
 \item The auctioneer observes excess demands, and adjusts the prices using (\ref{eq:AuctionO}).
 \end{itemize}
        }
\UNTIL {Market clears}.
\STATE Trades occur (Agreements are final).
\end{algorithmic}
\end{algorithm}
\begin{proposition}
\label{pr:WalConver}
Let $\left[\bar{\mathbf{x}}_{n}^{(s)},\mathbf{p}^{(s)}\right]_{s \in \mathcal{S}}$ be a Walrasian equilibrium under uncertainty. Then, in our setting, the Walras' tatonnement process with price adjustment rule (\ref{eq:AuctionO}) converges to $\left[\bar{\mathbf{x}}_{n}^{(s)},
\mathbf{p}^{(s)}\right]_{s \in \mathcal{S}}$ as $t \to \infty$.
\end{proposition}
\begin{IEEEproof}
See Section \ref{subSec:ThCon}.
\end{IEEEproof}
The process described in \textbf{Algorithm \ref{Alg:Auction}} requires a coordinator; however, in the absence of such a coordinator, the auction process can be implemented in a fully distributed manner. In doing so, the initial price and the price adjustment factor is known to all systems. Then every system (i) announces its demands, (ii) updates the excess demand based on the announced demands of all other bidders as well as its own, and (iii) updates the prices based on the excess demand. The process continues until convergence. 

Note that in engineering applications, for example in our system model when CPSs communicate over the wireless medium, the auction process does not cause heavy overhead, due to the following reason: Since the demand is not considered as private information, it can be simply announced through a control channel. Due to the broadcast nature of wireless medium, the signals are heard by all systems; in other words, no pairwise communication is required. After the convergence of the process, the channel becomes free. Usually, the convergence is fast, implying that the channel is not occupied for a long time. As conventional, we assume the existence of a noiseless control channel, implying that the signals arrive error-free at CPSs and/or the auctioneer. Such an assumption is justified by the existence of coding schemes that reduce the transmission error probability to almost zero \cite{Maghsudi15:CSE}; therefore, it is widely used in game theory and networks literature (see \cite{Maghsudi17:DUAEH} and \cite{Mochaourab15:DCA}, among many others). While the analysis of the problem with noisy communication channel is beyond the scope of our work, there are some research works that study such setting. For example, in 
\cite{Palguna16:SSA}, the authors assume that the channel from bidders to the auctioneer is noisy, whereas the reverse channel is noise-free. In this setting, they investigate a spectrum auction problem.  
\subsection{Complexity and Convergence Speed}
\label{sec:Complex}
For divisible goods, the convergence of the tatonnement process given by (\ref{eq:AuctionO}) is guaranteed only asymptotically, similar to many other tatonnement procedures. Thus, most often the auction is implemented to stop when the excess demand is only almost-zero, resulting in an approximate equilibrium. Moreover, the convergence speed of the Walrasian auction depends mainly on the price adjustment factor $\alpha$ as well as agents' utility functions, and thus cannot be determined rigorously. It is however important to note that $\alpha$ being too small slows down the convergence dramatically, whereas selecting $\alpha$ too large prevents the process from convergence.  

On the agents' side, the problem of calculating demands given the prices (budget) is a non-linear contentious knapsack problem with the number of inputs being the number of tasks $M$ at every state. For such problem polynomial-time algorithm exists \cite{Hochbaum07:CAN}. On the auctioneer side, calculating the excess demand and adjusting the price is linear in the number of CPSs $N$.
\section{Numerical Results}
\label{sec:NumG}
The numerical evaluation consists of two parts. In the first part, we implement the application scenario discussed in Section 
\ref{subsec:Examples}. We provide some numerical results to describe the applicability of the model and solution in an intuitive manner. In the second part, we consider a general example to demonstrate the theoretical results also by numerical analysis. The evolution of price and demand, the convergence rate, the incentive to form the grand coalition and other theoretical issues are discussed as well.
\subsection{A Toy Example}
\label{sec:NumToy}
Recall the application model we described in Section \ref{subsec:Examples}. There exist two SBSs in the network, i.e., $N=2$. Moreover, let 
$\alpha_{1}=0.9$ and $\alpha_{2}=0.7$. In addition, there are two possible states with respect to the weather, namely $\mathcal{S}=
\left\{Sunny, Windy \right\}$; In other words, the weather can be either \textit{sunny} or \textit{windy}. For the simplicity of notation, we label the sunny and the windy weather as State 1 and State 2, respectively. SBS 1 is provided with a number of solar cells as energy harvesting units. SBS 2 has a lower number of solar cells compared to SBS 1; instead, it has access to few wind-turbines. While a sunny weather is ideal to utilize solar cells for energy harvesting, it is not productive for wind-turbines. Consequently, we select 
$\rho_{12}^{(1)}=0.9$, $\rho_{12}^{(2)}=0.1$, $\rho_{22}^{(1)}=0.4$, $\rho_{22}^{(2)}=0.6$. For simplicity, we let $\lambda_{nm}^{(s)}=1$ for $n,m,s \in \left\{1,2\right\}$. The probability distributions assigned by the SBSs to the weather are selected randomly as 
$\mathbf{a}_{1}=(0.20,0.80)$ and $\mathbf{a}_{2}=(0.40,0.60)$. 

\begin{figure}[ht]
\centering
\includegraphics[width=0.38\textwidth]{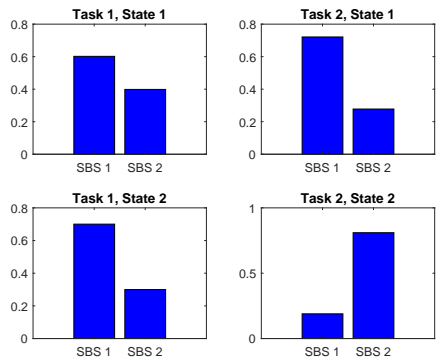}
\caption{Allocation of the computation and transmission tasks in State 1 and State 2.}
\label{Fig:TA}
\end{figure}

In \textbf{Fig. \ref{Fig:TA}}, we depict the task allocation resulted by the proposed method. It can be seen that in both states, a large fraction of transmission is allocated to SBS 1. Intuitively, such allocation is justified since SBS 1 is more likely to have a free channel to perform the relaying compared to SBS 2. Moreover, in State 1, i.e., if the weather is sunny, SBS 1 receives the larger part of the computation task as it has a larger number of solar cells, thereby more energy. In State 2, however, SBS 1 can afford only a low energy resource to perform the computation, due to the lack of sun; as a result, a part of the task is transferred to SBS 2, which can harvest the energy also from the wind. Therefore, in State 2, SBS 2 performs a large part of the computation task. Thus, the allocation of the computation tasks is along with the interests of SBSs, although it is performed before the weather is known. Hence, the SBSs are unlikely to deviate from it when the state is observed at a later point. It should be noted that, unlike the allocation profile of the computation task, the allocation profile of the transmission task is affected by the weather condition only slightly. This is due to the fact that, as described in Section \ref{subsec:Examples}, the required transmission power is supplied by the battery. The slight changes assure appropriate load-balancing and fairness in utility.
\subsection{A General Example}
\label{sec:Num}
We consider a CPS network consisting of four CPSs ($N=4$), three task types ($M=3$), and three states ($S=3$). The CPSs' performance indexes for carrying out tasks in different states are summarized in Table \ref{Tb:PerfIndex}. The values are generated simply at random. For every CPS $n \in \mathcal{N}$, the initial arrival of any task $m \in \mathcal{M}$ at every state $s \in \mathcal{S}$ is a random variable following an exponential distribution with parameter (rate) $\lambda_{nm}^{(s)}=n$. The probability distributions assigned by CPSs to states are selected randomly from the 3-dimensional probability space, and follow as $\mathbf{a}_{1}=(0.10,0.30,0.60)$, $\mathbf{a}_{2}=(0.20,0.50,
0.30)$, $\mathbf{a}_{3}=(0.34,0.33,0.33)$, and $\mathbf{a}_{4}=(0.90,0.05,0.05)$. The price adjustment factor is selected as $\alpha=0.01$. 
\begin{table}[ht]
\small
\centering
\caption{Performance index of every CPS $n \in \mathcal{N}$ for every task $m \in \mathcal{M}$ at every state $s \in \mathcal{S}$}
\subtable[State 1]{
  \begin{tabular}{ |c|c|c|c|c|}
  \hline
   Task &     CPS 1 & CPS 2 & CPS 3 & CPS 4           \\ \hline 
   1    &     0.80  & 0.34  & 0.72  & 0.42            \\ \hline
   2    &     0.21  & 0.68  & 0.22  & 0.78            \\ \hline
   3    &     0.26  & 0.19  & 0.65  & 0.71            \\ \hline
  \end{tabular}
  \label{Tb:StateOne}
}
\quad
\subtable[State 2]{
  \begin{tabular}{ |c|c|c|c|c|}
  \hline
   Task &     CPS 1 & CPS 2 & CPS 3 & CPS 4           \\ \hline 
   1    &     0.90  & 0.70  & 0.74  & 0.90            \\ \hline
   2    &     0.89  & 0.19  & 0.50  & 0.61            \\ \hline
   3    &     0.33  & 0.20  & 0.48  & 0.62            \\ \hline
  \end{tabular}
  \label{Tb:StateTwo}
}
\quad
\subtable[State 3]{
  \begin{tabular}{ |c|c|c|c|c|}
  \hline
   Task &     CPS 1 & CPS 2 & CPS 3 & CPS 4           \\ \hline 
   1    &     0.86  & 0.21  & 0.19  & 0.98            \\ \hline
   2    &     0.81  & 0.24  & 0.49  & 0.71            \\ \hline
   3    &     0.88  & 0.89  & 0.21  & 0.50            \\ \hline
  \end{tabular}
  \label{Tb:StateThree}
}
\label{Tb:PerfIndex}
\end{table}
%

\begin{figure}[ht]
\centering
\includegraphics[width=0.35\textwidth]{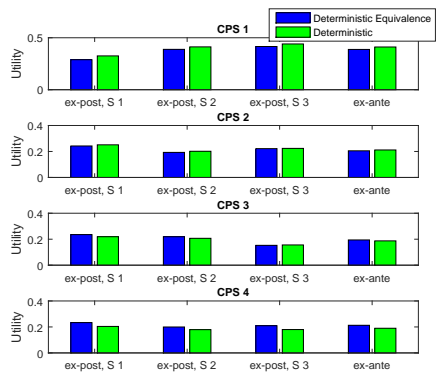}
\caption{The utility resulted from deterministic equivalence allocation for the stochastic problem, compared to the deterministic game.}
\label{Fig:DetEqu}
\end{figure}

First we investigate the effect of using deterministic equivalence in the analysis. We initially consider stochastic task arrival, where CPSs apply the proposed approach to agree on task allocation. In \textbf{Fig. \ref{Fig:DetEqu}}, we show the ex-ante and ex-post utility of each CPS $n \in \mathcal{N}$, normalized by the aggregate utility, i.e., $\frac{u_{n}}{\sum_{n\in \mathcal{N}}u_{n}}$. In essence, this is the utility achieved by the task allocation resulted from using the concept of deterministic equivalence (i.e., dictated by 
$\left[\mathbf{r}_{n}^{(s)}\right]_{n\in \mathcal{N}, s\in \mathcal{S}}$). In another experiment, we assume that the tasks arrived at every CPS are deterministic and known. We perform only the conventional auction process, and show the utility in \textbf{Fig. 
\ref{Fig:DetEqu}}, as described before. From this figure, it can be concluded that by using the deterministic equivalence, the effect of stochastic task arrival is almost eliminated; that is, the users are indifferent between receiving the deterministic equivalent payoff or the stochastic payoff, as expected. 

\begin{figure}[ht]
\centering
\includegraphics[width=0.48\textwidth]{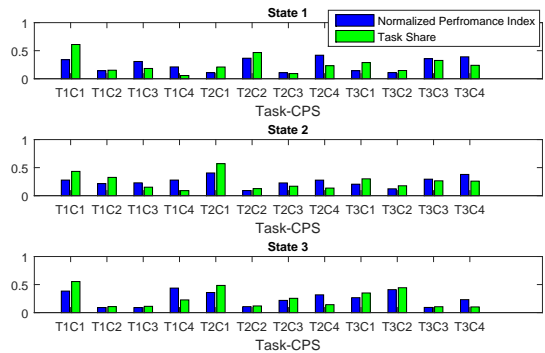}
\caption{The relation between efficiency (relative performance index) and allocated task.}
\label{Fig:EffAll}
\end{figure}

Moreover, in \textbf{Fig. \ref{Fig:EffAll}}, we illustrate the relation between the performance index and task allocation. As expected, a CPS with high performance index for some task is more likely to receive larger share of that task in the final allocation, compared to a CPS with a low performance index for that task. This results in higher utility for CPSs, and also improves the overall performance of the platform.     

\begin{figure}[ht]
\centering
\includegraphics[width=0.38\textwidth]{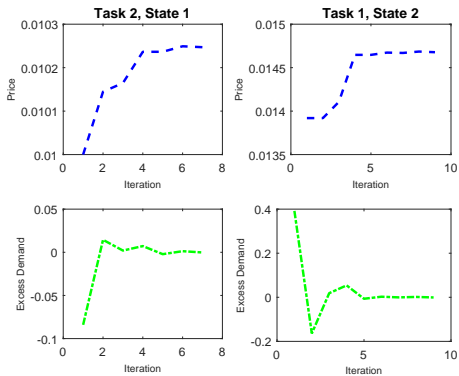}
\caption{The changes in price and excess demand for two exemplary state-contingent tasks.}
\label{Fig:PrDe}
\end{figure}

\begin{figure}[ht]
\centering
\includegraphics[width=0.26\textwidth]{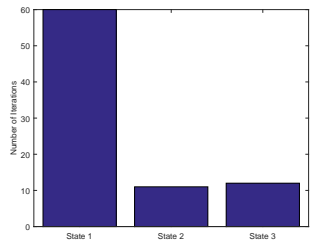}
\caption{Required number of iterations to converge (market clearing).}
\label{Fig:Conv}
\end{figure}

\textbf{Fig. \ref{Fig:PrDe}} shows the relation between the excess demand and the price of two exemplary state contingent tasks. It can be seen that, as expected, the excess demand reduces as the price increases. The entire agreement process ends when the market clears for all state contingent tasks, i.e., $\sum_{n \in \mathcal{N}}\mathbf{r}_{n}^{(s)}=\mathbf{1}$ for all $s \in \mathcal{S}$. The number of iterations required for the market clearing for every state contingent task is shown in \textbf{Fig. \ref{Fig:Conv}}. Note that as described in Section \ref{sec:Complex}, we implement the tatonnement process to stop when the excess demand is very small (here $0.01$). From the figure, the entire allocation process converges in $60$ iterations. Note that as we discussed in Section \ref{sec:Complex}, the convergence speed cannot be determined rigorously, and may vary based on many factors such as the price adjustment coefficient, random initial task loads, performance indexes, etc.   

Next, we show that the task allocation belongs to the strong sequential core; that is, joining the grand coalition is beneficial for every CPS $n \in \mathcal{N}$ ex-ante, and the ex-ante allocation lies in the core of the ex-post games. Since the efficiency prerequisite is satisfied by the market clearing condition (every task is allocated in full), we only need to check the coalitional deviations prerequisites. This means that there should be no coalition in which at least one member is better off leaving the grand coalition while all other members remain indifferent. 

\begin{figure}[ht]
\centering
\caption{Ex-ante and ex-post coalitional rationality.}
\label{Fig:core}
\subfigure[Ex-Ante]{
\centering 
\includegraphics[width=0.43\textwidth]{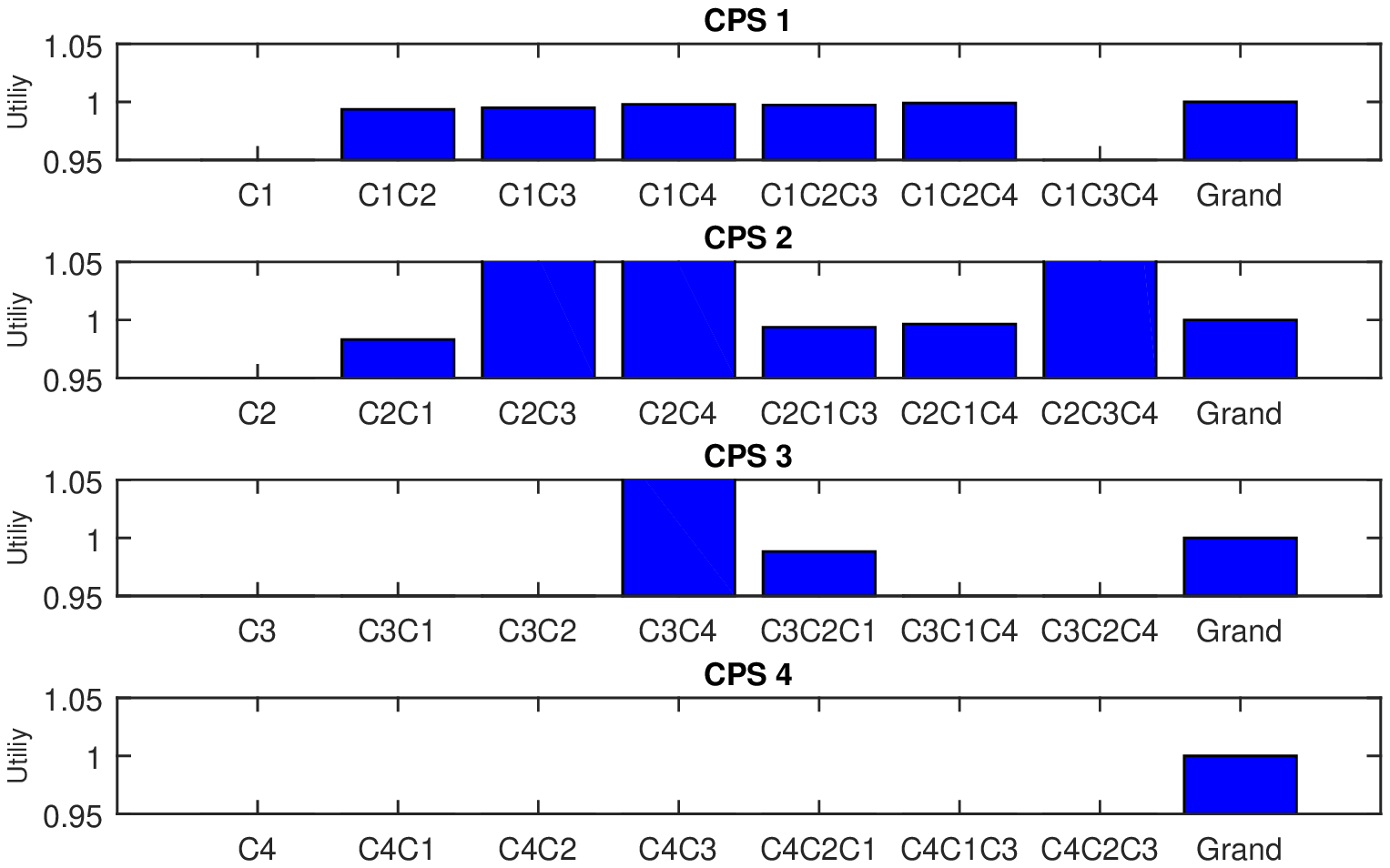}
\label{Fig:coreAnte}
}
\quad
\subfigure[Ex-post, State 1]{
   \centering
  \includegraphics[width=0.43\textwidth]{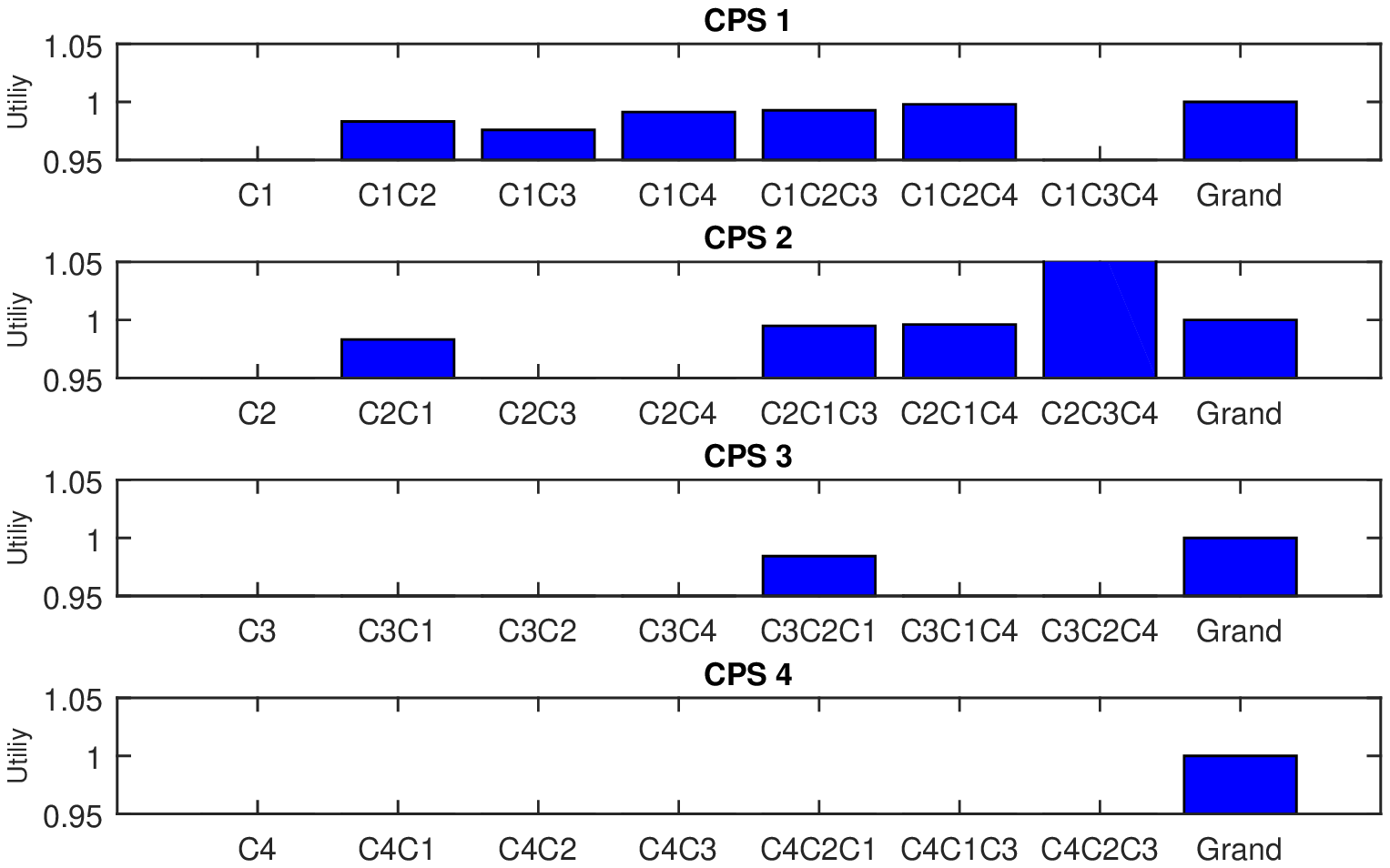}
  \label{Fig:corePostA}
}
\quad
\subfigure[Ex-post, State 2]{
   \centering
  \includegraphics[width=0.43\textwidth]{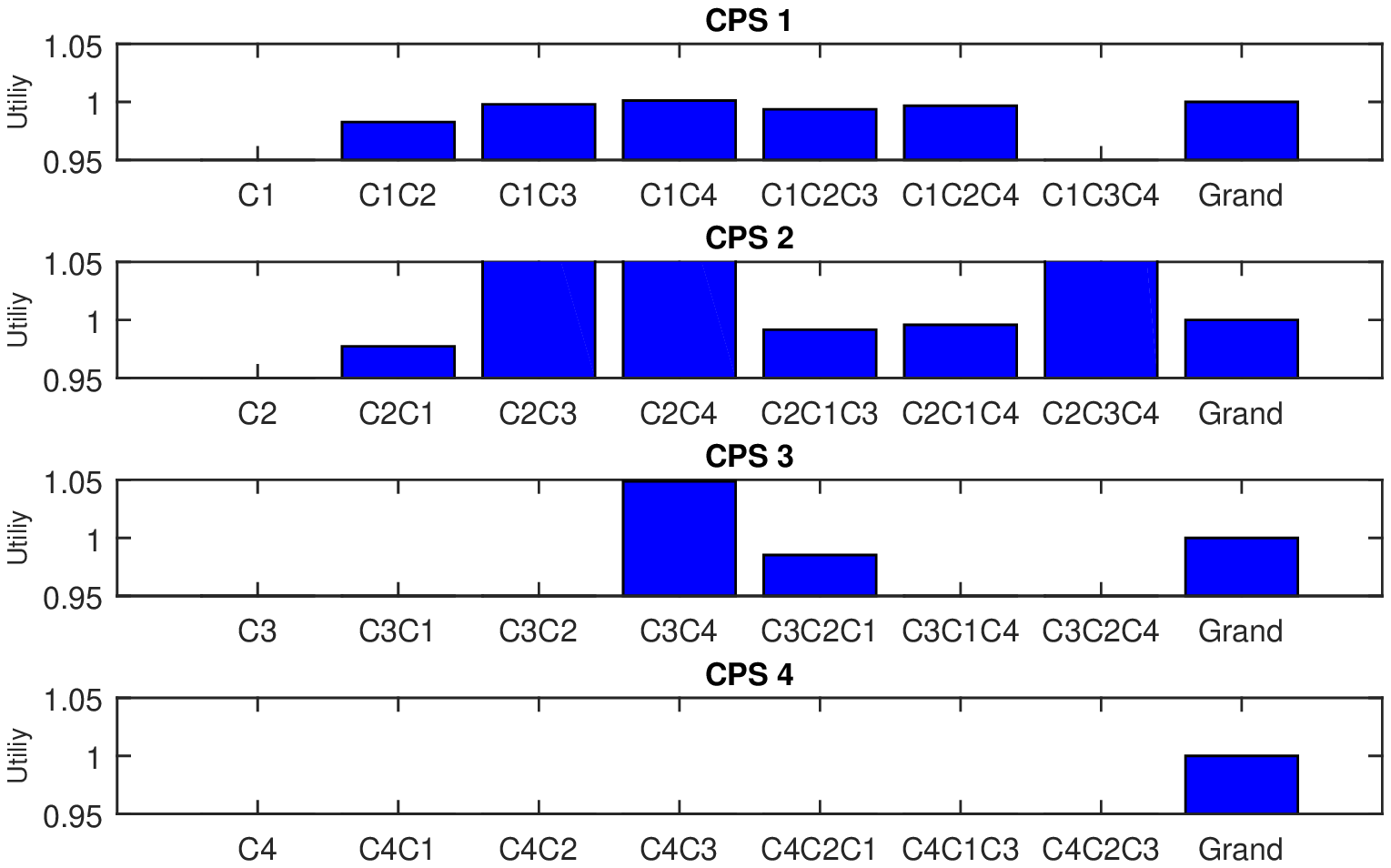}
  \label{Fig:corePostB}
}
\quad
\subfigure[Ex-post, State 3]{
   \centering
  \includegraphics[width=0.43\textwidth]{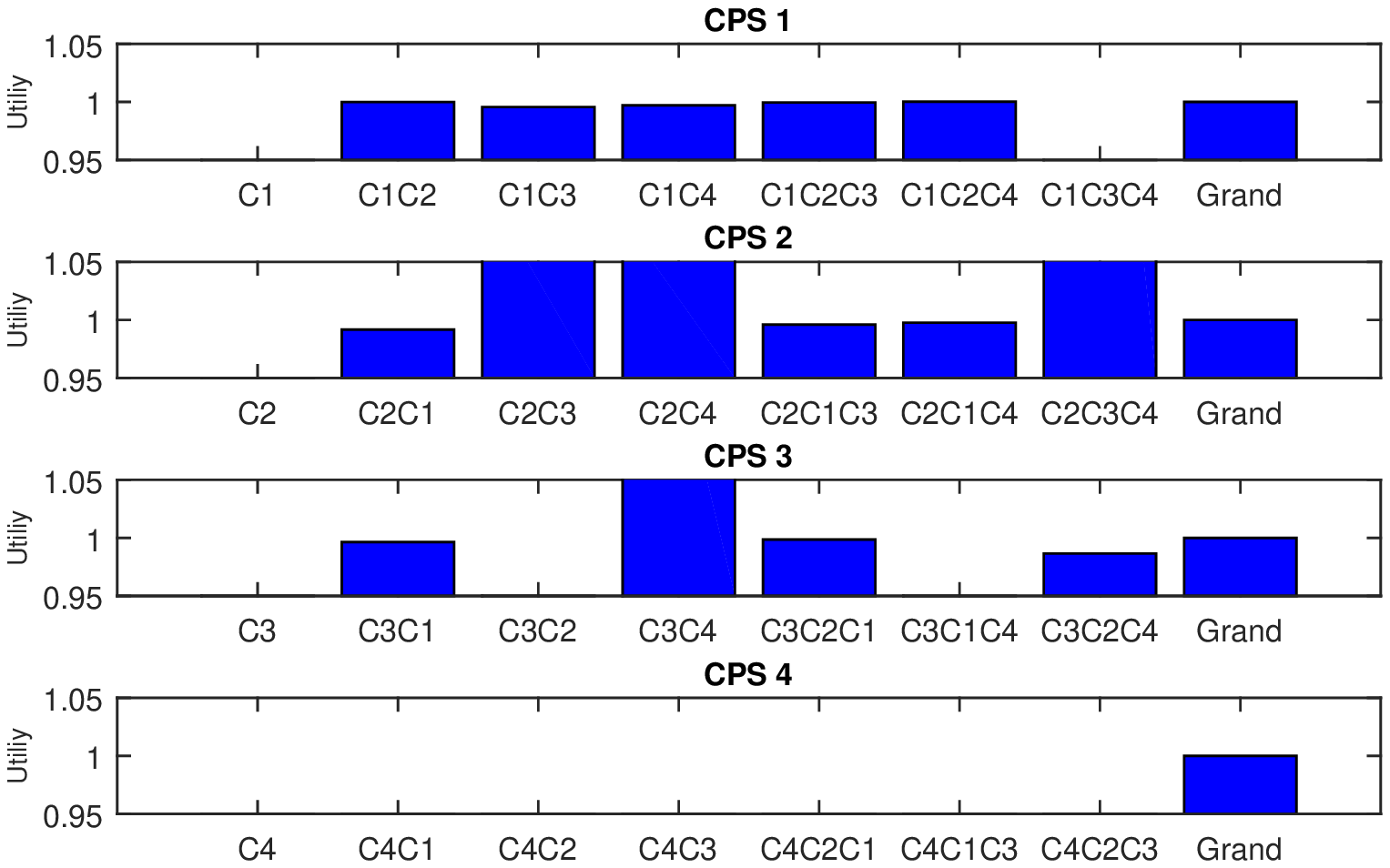}
  \label{Fig:corePostC}
}
\end{figure}

To perform the experiment, we first simulate some task arrival according to the described exponential distribution. In 
Figs. \ref{Fig:core}(a)-\ref{Fig:core}(d), we illustrate the maximum ex-ante and ex-post achievable reward of every CPS at every possible coalition, including the grand coalition. For each CPS, all values are normalized by its utility resulted from our proposed task allocation scheme, i.e., as dictated by $\left[\mathbf{r}_{n}^{(s)}\right]_{n \in \mathcal{N}, s\in \mathcal{S}}$. Note that $\left[\mathbf{r}_{n}^{(s)}\right]_{n \in \mathcal{N}, s\in \mathcal{S}}$ is agreed upon before the task arrival. The values which do not appear in the diagrams are less than 95\% of the reward of grand coalition, and therefore cannot be observed. \textbf{Fig. \ref{Fig:core}(a)} shows the normalized expected utility of every CPS ex-ante. It can be observed that for the grand coalition, the ratio is always equal to one. This means that our allocation scheme achieves the best performance, despite being performed prior to the actual job arrival. Moreover, there is no coalitional deviation (including singleton coalitions) in which at least one member benefits, while other members remain indifferent. Thus the allocation is stable. In Figs. \textbf{\ref{Fig:core}(b)}, \textbf{\ref{Fig:core}(c)}, and \textbf{\ref{Fig:core}(d)}, we assume that state 1, state 2, and state 3 are realized, respectively. The normalized utility for the ex-post games are then depicted. The figures show that the grand coalition is also ex-post stable, i.e., no coalition of CPSs benefits from a deviation even after the realized state is revealed. Together with the efficiency condition, we can conclude that the allocation dictated by $\left[\mathbf{r}_{n}^{(s)}\right]_{n \in \mathcal{N}, s\in \mathcal{S}}$ cannot be improved upon by any other allocation ex-ante and also ex-post. Therefore it lies in the strong sequential core.     

\begin{figure}[H]
\centering
\includegraphics[width=0.23\textwidth]{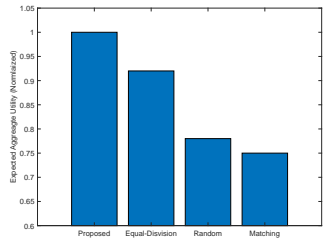}
\caption{Performance Comparison.}
\label{Fig:Comparison}
\end{figure}

Finally, we evaluate the performance of our approach in terms of the expected social welfare (expected aggregate utility), by comparing it with the following allocation methods: 
\begin{itemize}
\item Weighted Matching: We construct a bipartite graph consisting of CPSs on the one side and tasks on the other side. The performance indexes (types) of CPSs are used as the edges' weights. By using a bipartite maximum matching algorithm \cite{Maghsudi17:DUA}, each task is allocated to only one CPS, in a way that the sum of the weights of assignment edges is maximized.
\item Random Allocation: The tasks are divided between the CPSs randomly;
\item Equal Allocation: Each task is equally divided between the CPSs.
\end{itemize}
The results are depicted \textbf{Fig. \ref{Fig:Comparison}}, which shows the superior performance of our proposed method. Naturally, the achieved gain is not fixed and might vary depending on many parameters such as CPSs' types. 
\section{Conclusion}
\label{sec:Conc}
We considered multi-state stochastic cooperative games, where (i) the reward of every coalition is a state-dependent random variable and 
(ii) the nature's state is random. The players, characterized by state-dependent utility functions, have to divide the stochastic reward, before the state is realized. We applied the strong sequential core as a desirable solution concept. We then solved the problem by using the concept of deterministic equivalence and expected utility. For an exemplary application, namely task allocation in cyber-physical systems, we established the non-emptiness of core and characterized it with respect to uniqueness and optimality. We proved that in our setting, the Walrasian equilibrium under state uncertainty lies in the strong sequential core, and thus the core can be implemented by using the Walras' auction. Numerical results confirm the theoretical analysis and showed the applicability of the proposed model and solution. 
\section{Appendix}
\label{sec:App}
Before proceeding to proofs, we state some auxiliary definitions and results, which were left out from the main text in order to maintain the readability and consistency.
\subsection{Technical Preliminaries for Cooperative Games}
\label{subsec:TechPre}
\begin{definition}[Feasible Allocation]
\label{de:FeasAll}
Let some allocation $\bar{\mathbf{x}}$ be given. The allocation $\mathbf{x}_{c}=\left[\mathbf{x}_{n,c}^{(s)}\right]_{s \in \mathcal{S}, n 
\in \mathcal{N}}$ is feasible for a coalition $c \in \mathcal{C}$ at state $s \in \mathcal{S}$ if $\mathbf{x}_{c}^{-(s)}=\bar{\mathbf{x}}_{c}^{-(s)}$ and $\sum_{n \in c}\mathbf{x}_{n,c}^{(s)}\leq \mathbf{w}_{c}^{(s)}$, where $\mathbf{w}_{c}^{(s)}$ is the value of the coalition at state $s \in \mathcal{S}$. In words: (i) The members of the coalition take allocations outside $s \in \mathcal{S}$ as given; (ii) In every state $s \in \mathcal{S}$, the members of a coalition can redistribute at most their wealth. For an allocation to be feasible at stage $0$ 
(before knowing the state), the second condition must hold for every state $s \in \mathcal{S}$.
\end{definition}
\begin{definition}[Deviation]
\label{de:Dev}
Let some allocation $\bar{\mathbf{x}}$ be given. A coalition $c \in \mathcal{C}$ can deviate from $\bar{\mathbf{x}}$ at state $s \in 
\mathcal{S}$ if there exists a feasible allocation $\mathbf{x}_{c}^{(s)}$ for $c$ at state $s$ such that $u_{n,c}\left(\mathbf{x}_{n,c}
\right)>u_{n,c}\left(\bar{\mathbf{x}}_{n,c}\right)$ for all $n \in \mathcal{C}$. Such allocations are called deviations, which are not necessarily self-enforcing. 
\end{definition}
\subsection{Technical Preliminaries for Exchange Economy}
\label{subsec:TechPreEx}
%
\begin{definition}[Feasible Exchange]
\label{de:FAllocation}
In an exchange economy $\Omega$, an allocation $\bar{\mathbf{x}}$ is a feasible exchange if $\sum_{n \in \mathcal{N}} \bar{\mathbf{x}}_{n}\leq \sum_{n \in \mathcal{N}}\mathbf{q}_{n}$. 
\end{definition}
\begin{definition}[Pareto-Efficient Exchange]
\label{de:PrOptimal}
In an exchange economy $\Omega$, a feasible exchange $\bar{\mathbf{x}}$ is Pareto-efficient if there is no other feasible exchange 
$\mathbf{x}$ such that $u_{n}(\mathbf{x}_{n}) \geq u_{n}(\bar{\mathbf{x}}_{n})$ for all $n \in \mathcal{N}$ with strict inequality for some 
$n \in \mathcal{N}$.
\end{definition}
\begin{definition}[Social-Optimal Exchange]
\label{de:SoOptimal}
A feasible exchange profile $\bar{\mathbf{x}}^{(s)}$, $s \in \mathcal{S}$, is social-optimal if it maximizes the aggregate agents' utility 
(social welfare). Formally, any feasible solution to the following optimization problem is a social-optimal exchange. 
\begin{equation}
\label{eq:OptPrT}
\underset{\mathbf{x}^{(s)} \in \mathcal{X}^{(s)}}{\textup{maximize}}~\sum_{n \in \mathcal{N}}u_{n}\left(\left[\mathbf{x}^{(s)}\right]_{s \in 
\mathcal{S}}\right).
\end{equation}
\end{definition}
\begin{definition}[Local Non-Satiated]
\label{de:LoSatis}
If $\mathcal{X}$ is a consumption set, then for any $\mathbf{x} \in \mathcal{X}$ and every $\epsilon>0$, there exists a $\mathbf{y} \in \mathcal{X}$ such that $\left\|\mathbf{y}-\mathbf{x} \right\|\leq \epsilon$ and $\mathbf{y}$ is preferred to $\mathbf{x}$. 
\end{definition}
\begin{definition}[Gross Substitutes \cite{Levin06:GE}]
\label{de:GS}
A demand function $x_{n}$ satisfies the gross substitutes property if, for any two price vectors $\mathbf{p}$ and $\mathbf{p}'$ such that 
$p'_{k} \geq p_{k}$ and $p'_{l} \geq p_{l}$ for all $l \neq k$, then $x_{l}(\mathbf{p}')>x_{l}(\mathbf{p})$ for all $l \neq k$.
\end{definition}
\begin{lemma}[\hspace{1sp}\cite{Gul99:WEGS}]
\label{Lm:GSAC}
Additive concave and separable additive utility functions satisfy the gross substitutes property.
\end{lemma}
\begin{lemma}[\hspace{1sp}\cite{Levin06:GE}]
\label{Lm:ExcDem}
If each individual has a utility function satisfying the gross substitutes property, then both the individual and aggregate excess demand functions satisfy the gross substitutes property as well.
\end{lemma}
\subsection{Technical Preliminaries for Walrasian Equilibrium}
\label{subsec:TechPreWa}
In this section we describe some results regarding the existence and optimality of Walrasian (Arrow-Debreu) equilibrium. Note that the following theorems were originally established for competitive markets without uncertainty (single-state), but were later shown in 
\cite{Debreu87:TTV} to also hold for competitive market with \textit{uncertainty}.
\begin{theorem}[Existence of Walrasian Equilibrium \cite{Colell85:MT}]
\label{Th:Ex}
Walrasian equilibrium $\left[\mathbf{x}_{n}^{(s)},\mathbf{p}^{(s)}\right]_{s \in \mathcal{S}}$ exists for an exchange economy with divisible goods if (i) for every agent, the utility function is continuous, increasing, concave; and (ii) $\mathbf{q}_{n}^{(s)}>0$ for all $s \in \mathcal{S}$ and $n \in \mathcal{N}$.  
\end{theorem}
\begin{theorem}[Efficiency of Walrasian Equilibrium \cite{Gul99:WEGS}]
\label{Th:Opt}
Let $\left[\mathbf{x}_{n}^{(s)},\mathbf{p}^{(s)}\right]_{s \in \mathcal{S}}$ be a Walrasian equilibrium. Then it is Pareto-efficient for an exchange economy with divisible goods if the utility function is increasing. 
\end{theorem}
\begin{theorem}[Uniqueness of Walrasian Equilibrium \cite{Levin06:GE}]
\label{Th:Uniq}
For an exchange economy $\Omega$ with divisible goods, if the aggregate excess demand function $z(\cdot)$ satisfies the gross substitutes property, then the economy has at most one Walrasian equilibrium, i.e., $z\left(\mathbf{p}\right)=0$ has at most one 
(normalized) solution. 
\end{theorem}
\begin{theorem}[Convergence of Walras' Tatonnement \cite{Levin06:GE}]
\label{Th:Conv}
Consider an exchange economy $\Omega$ and suppose that $\left[\mathbf{p}^{(s)}\right]_{s \in \mathcal{S}}$ is a Walrasian equilibrium price vector. Suppose that the commodities are divisible and the aggregate excess demand function $z(\cdot)$ satisfies the gross substitutes property. Then the tatonnement process with price adjustment rule (\ref{eq:AuctionO}) converges to the relative prices of 
$\left[\mathbf{p}^{(s)}\right]_{s \in \mathcal{S}}$ as $t \to \infty$ for any initial condition $\mathbf{p}(t=0)$. 
\end{theorem}
\subsection{Proof of Proposition \ref{pr:UtiAss}}
\label{SubSec:Utility}
Recall that by (\ref{eq:MachUtil}), we have $u_{nm}^{(s)}\left(x_{nm}^{(s)}\right)=\rho_{nm}^{(s)}\left(1-e^{-\frac{1}{\rho_{nm}^{(s)}}
x_{nm}^{(s)}}\right)$. Therefore, 
\begin{itemize}
\item The continuity of the function (\ref{eq:UtilityOne}) follows by the continuity of the exponential function 
$e^{-\frac{1}{\rho_{nm}^{(s)}}x_{nm}^{(s)}}$ for all $m \in \mathcal{M}$. The continuity of the exponential function is a standard result proved by a straightforward application of the definition of continuity. Since the result exists in standard mathematics text books, we do not state it here to save the space.  
\item The first derivation of (\ref{eq:UtilityOne}) yields $u\myprime_{nm}^{(s)}\left(x_{nm}^{(s)}\right)=\sum_{m\in \mathcal{M}} 
e^{-\frac{1}{\rho_{nm}^{(s)}}x_{nm}^{(s)}}$, which is positive for $x_{nm}^{(s)}>0$, implying that the function is monotone increasing. 
\item The second derivative is given by $u\mydprime_{nm}^{(s)}\left(x_{nm}^{(s)}\right)=\sum_{m\in \mathcal{M}}\frac{-1}{\rho_{nm}^{(s)}}
e^{-\frac{1}{\rho_{nm}^{(s)}}x_{nm}^{(s)}}$, which is negative, meaning that the function is concave. 
\end{itemize}
The second part of the proposition simply follows by the definition of $u_{n}\left(\mathbf{x}_{n}\right)$.
\subsection{Proof of Proposition \ref{pr:exponential}}
\label{SubSec:exponential}
We follow the line suggested in \cite{Suijs99:SCG}. For simplicity of notation, define $\alpha_{1}=\rho_{nm}^{(s)}$, $\alpha_{2}=
-\rho_{nm}^{(s)}$, and $b=\frac{-1}{\rho_{nm}^{(s)}}$. Recall that the utility function is given by $u(\mathbf{x})=\alpha_{1}+\alpha_{2}
e^{b\mathbf{x}}$. Moreover, for $\mathbf{x}, \mathbf{y} \in \mathcal{R}_{>0}$, $\mathbf{x} \geq \mathbf{y}$ if $\mathbb{E}
\left[\mathbf{x}\right] \geq \mathbb{E}\left[\mathbf{y}\right]$. It is obvious that
\begin{equation}
\label{eq:expected}
\mathbb{E}\left[u(\mathbf{x})\right]=\alpha_{1}+\alpha_{2}\mathbb{E}\left[e^{b\mathbf{x}} \right].
\end{equation}
We define \cite{Suijs99:SCG} 
\begin{equation}
\label{eq:m}
\mathbf{d}(\mathbf{x})=u^{-1}\left(\mathbb{E}\left[u(\mathbf{x})\right]\right).
\end{equation}
Clearly, 
\begin{equation}
\label{eq:u}
u^{-1}(\boldsymbol{\tau})=\frac{1}{b}\ln \left(\frac{\boldsymbol{\tau}-\alpha_{1}}{\alpha_{2}}\right).
\end{equation}
By (\ref{eq:m}) and (\ref{eq:u}), we have
\begin{equation}
\label{eq:mTwo}
\begin{aligned} 
\mathbf{d}(\mathbf{x})&=\frac{1}{b}\ln \left(\frac{\mathbb{E}\left[u(\mathbf{x})\right]-\alpha_{1}}{\alpha_{2}}\right)\\
&=\frac{1}{b}\ln \left(\frac{\alpha_{1}+\alpha_{2}\mathbb{E}\left[e^{b\mathbf{x}}\right]-\alpha_{1}}{\alpha_{2}}\right)\\
&=\frac{1}{b}\ln \left(\mathbb{E}\left[e^{b\mathbf{x}}\right]\right).
\end{aligned}
\end{equation}
Consequently,
\begin{equation}
\label{eq:mThree}
\begin{aligned}
u\left(\mathbf{d}(\mathbf{x})\right)&=\alpha_{1}+\alpha_{2}e^{b\mathbf{d}(\mathbf{x})}\\ 
 &=\alpha_{1}+\alpha_{2}\mathbb{E}\left[e^{b\mathbf{x}}\right]
\end{aligned}
\end{equation}
where the second inequality simply follows by using (\ref{eq:mTwo}). In order to prove that the $\mathbf{d}(\mathbf{x})$ given by 
(\ref{eq:mTwo}) is the certainty equivalent of $x$, in what follows we show that all requirements stated in Definition 
\ref{de:DetEqui} are satisfied. 
\begin{enumerate}[(a)]
\item By (\ref{eq:expected}) and (\ref{eq:mThree}), we see that $u\left(\mathbf{d}(\mathbf{x})\right)=\mathbb{E}\left[u(\mathbf{x})\right]$. Thus, the player is indifferent between receiving the stochastic reward $\mathbf{x}$ or the deterministic reward $\mathbf{d}(\mathbf{x})$, so that $\mathbf{x} \approx \mathbf{d}(\mathbf{x})$. Thus the first condition holds.
\item The \textit{Von Neumann-Morgenstern} preference relation is stated in (\ref{eq:preference}). Consequently, 
\begin{equation}
\label{eq:pre}
\mathbf{x} \geqslant_{n} \mathbf{y}~~\text{if}~~\alpha_{1}+\alpha_{2}\mathbb{E}\left[e^{b\mathbf{x}}\right]>
\alpha_{1}+\alpha_{2}\mathbb{E}\left[e^{b\mathbf{y}}\right].
\end{equation}
Thus
\begin{equation}
\label{eq:preT}
\ln \left(\mathbb{E}\left[e^{b\mathbf{x}}\right]\right)\geq \ln \left(\mathbb{E}\left[e^{b\mathbf{y}}\right]\right),
\end{equation}
which, by (\ref{eq:mTwo}), implies $\mathbf{d}(\mathbf{x})\geq \mathbf{d}(\mathbf{y})$.
The reverse follows similarly, hence condition two holds.   
\item The third condition holds since by using (\ref{eq:mTwo}), for deterministic $\mathbf{k}$ we have
\begin{equation}
\label{eq:fixed}
\mathbf{d}(\mathbf{k})=\frac{1}{b}\ln \left(\mathbb{E}\left[e^{b\mathbf{k}}\right]\right)=\mathbf{k}.
\end{equation}
\item By using (\ref{eq:mTwo}) we have
\begin{equation}
\label{eq:zero}
\begin{aligned} 
\mathbf{d}\left(\mathbf{x}-\mathbf{d}(\mathbf{x})\right)&=\frac{1}{b}\ln \left(\mathbb{E}\left[e^{b(\mathbf{x}-\mathbf{d}
(\mathbf{x}))}\right]\right)\\
&=\frac{1}{b}\ln \left(e^{-b\mathbf{d}(\mathbf{x})}\right)+\frac{1}{b}\ln \left(\mathbb{E}\left[e^{b\mathbf{x}}\right]\right)\\
&=-\mathbf{d}(\mathbf{x})+\mathbf{d}(\mathbf{x})=0.
\end{aligned}
\end{equation}
Therefore condition four holds.
\item Consider deterministic $\mathbf{k},\mathbf{k}' \in \mathcal{R}_{>0}^{M}$, with $\mathbf{k}<\mathbf{k}'$. Then
\begin{equation}
\label{eq:fixedTwo}
\ln \left(\mathbb{E}\left[e^{b\mathbf{k}}e^{b\mathbf{x}}\right]\right)\leq \ln \left(\mathbb{E}\left[e^{b\mathbf{k}'}
e^{b\mathbf{x}}\right]\right),
\end{equation}
which by (\ref{eq:mTwo}) implies $\mathbf{d}(\mathbf{x}+\mathbf{k})\leq \mathbf{d}(\mathbf{x}+\mathbf{k}')$. Thus condition five holds.
\end{enumerate} 
%
\subsection{Proof of Proposition \ref{Pr:WalChar}}
\label{SubSec:PropUtiExpN}
In our designed multi-state stochastic task exchange economy, for each CPS $n \in \mathcal{N}$ and in every state $s \in \mathcal{S}$, the utility of performing some task $m \in \mathcal{M}$ is given by (\ref{eq:MachUtil}). Moreover, for CPS $n \in \mathcal{N}$ at state $s\in \mathcal{S}$, the initial arrived load of type $m \in \mathcal{M}$ follows an exponential distribution with parameter 
$\lambda_{nm}^{(s)}$. Also, the total utility at state $s$, i.e., $u_{n}^{(s)}$, follows by (\ref{eq:UtilityOne}). 

To establish the proposition, first we need to calculate the deterministic equivalent. By (\ref{eq:utiexpDet}), the deterministic 
(equivalent) utility function of CPS $n \in \mathcal{N}$ is given as 
\begin{equation}
\label{eq:proofOne}
d_{nm}^{(s)}=-\rho_{nm}^{(s)}\ln\left(\mathbb{E} \left[e^{\frac{-1}{\rho_{nm}^{(s)}}\left(r_{nm}^{(s)}Q_{m}^{(s)}\right)}\right] \right)
\end{equation}
Thus we first need to solve for $\mathbb{E}\left[e^{\frac{-1}{\rho_{nm}^{(s)}}\left(r_{nm}^{(s)}Q_{m}^{(s)}\right)}\right]$. Due to 
(\ref{eq:Sum}), we have
\begin{equation}
\label{eq:proofTwo}
\begin{aligned}
\mathbb{E}\left[e^{\frac{-1}{\rho_{nm}^{(s)}}\left(r_{nm}^{(s)}Q_{m}^{(s)}\right)}\right]&=
\mathbb{E}\left[e^{\frac{-1}{\rho_{nm}^{(s)}}\left(r_{nm}^{(s)}\sum_{j\in \mathcal{N}}q_{jm}^{(s)}\right)}\right]\\
&=\prod_{j \in \mathcal{N}}{\mathbb{E}\left[e^{\frac{-1}{\rho_{nm}^{(s)}}r_{nm}^{(s)}q_{jm}^{(s)}}\right]},
\end{aligned}
\end{equation}
where the equality follows due to the independence of $q_{jm}^{(s)}$, $j \in \mathcal{N}$. Let $q_{jm}^{(s)}$ be distributed according to an exponential distribution with parameter $\lambda_{jm}^{(s)}$. Then 
\begin{equation}
\label{eq:proofThree}
\begin{aligned}
\mathbb{E}\left[e^{\frac{-1}{\rho_{nm}^{(s)}}r_{nm}^{(s)}q_{jm}^{(s)}}\right]&=\int_{0}^{\infty}e^{\frac{-1}{\rho_{nm}^{(s)}}r_{nm}^{(s)}}\lambda_{jm}^{(s)}e^{-\lambda_{jm}^{(s)}q_{jm}^{(s)}}dq\\
&=\frac{\lambda_{jm}^{(s)}}{\lambda_{jm}^{(s)}+\frac{1}{\rho_{nm}^{(s)}}r_{nm}^{(s)}}.
\end{aligned}
\end{equation}
By (\ref{eq:proofTwo}) and (\ref{eq:proofThree}), (\ref{eq:proofOne}) can be written as 
\begin{equation}
\label{eq:proofFour}
\begin{aligned}
d_{nm}^{(s)}&=\rho_{nm}^{(s)}\ln\left(\prod_{j \in \mathcal{N}} \frac{\lambda_{jm}^{(s)}+\frac{1}{\rho_{nm}^{(s)}}r_{nm}^{(s)}}
{\lambda_{jm}^{(s)}} \right)\\
&=\rho_{nm}^{(s)}\sum_{j \in \mathcal{N}}\ln \left(\frac{\lambda_{jm}^{(s)}+\frac{1}{\rho_{nm}^{(s)}}r_{nm}^{(s)}}{\lambda_{jm}^{(s)}} 
\right).
\end{aligned}
\end{equation}
Since the utility function defined by (\ref{eq:MachUtil}) is continuous and monotonically increasing, maximizing 
$u_{nm,D}^{(s)}\left(d_{nm}^{(s)}\right)$ is equivalent to maximizing $d_{nm}^{(s)}$. Moreover, the function defined by (\ref{eq:proofFour}) is continuous, monotonically increasing and concave. Thus, by (\ref{eq:DetEqTheoremTwo}), the total utility is the sum of continuous, monotonically increasing and concave functions, hence demonstrating the characteristics required by Theorem \ref{Th:Ex}. Moreover, by Assumption \ref{ass:InitV}, the requirement on $\mathbf{q}_{D}^{(s)}$ is satisfied as well. Hence, by Theorem \ref{Th:Ex}, Walrasian equilibrium exists, and is also Pareto-efficient due to Theorem \ref{Th:Opt}.

Since the utility function given by (\ref{eq:DetEqTheoremTwo}) is additive separable, it satisfies the gross substitutes property, according to Lemma \ref{Lm:GSAC}. Then, due to Lemma \ref{Lm:ExcDem}, the excess demand function also satisfies the gross substitutes property. The result therefore follows from Theorem \ref{Th:Uniq}. Moreover, by Definition \ref{de:PrOptimal} and Definition \ref{de:SoOptimal}, every social-optimal allocation is also Pareto-efficient. Consequently, since there exists only one Pareto-efficient allocation, it is social-optimal as well.
\subsection{Proof of Theorem \ref{th:WalCore}}
\label{SubSec:ProofThOne}
In order to proof this proposition, it suffices to establish that both axioms of Definition \ref{de:WeakCore} hold. 

By (\ref{eq:DetEqivThree}), we know that $v_{n,D}\left(\left[\mathbf{x}_{n}^{(s)}\right]_{s \in \mathcal{S}}\right)=\sum_{s=1}^{S}a^{(s)}
u_{n,D}^{(s)}\left(\mathbf{x}_{n}^{(s)}\right)$. Thus, Proposition \ref{pr:UtiAss} also holds for $v_{n,D}$; that is, it is state-separable, are concave and monotonically increasing. Consequently, given Assumption \ref{ass:InitV}, it follows from Theorem \ref{Th:Ex} that for the game $\mathfrak{G}_{D}\left(\mathcal{N},v_{n,D}\right)$, Walrasian equilibrium exists. 

Moreover, as established in the proof of Proposition \ref{Pr:WalChar}, the excess demand function (\ref{eq:Execss}) satisfies the gross substitutes property. Together with Theorem \ref{Th:Uniq}, this results in the uniqueness of Walrasian equilibrium. Finally, from Theorem \ref{Th:Opt} and the uniqueness of equilibrium, it follows that it is Pareto-efficient and social-optimal. As a result, no coalition (including the grand coalition) is able to improve by means of deviating and blocking the Walrasian equilibrium. In fact, due to Theorem 
\ref{th:LieCore}, the Walrasian equilibrium lies inside the core of the ex-ante cooperative game $\mathfrak{G}_{D}$ (before revealing the state uncertainty). Thus the second condition of Definition \ref{de:WeakCore} is satisfied. 

Now we prove that the first condition is satisfied as well. That is, the Walrasian equilibrium under uncertainty belongs to the core of the state game $\mathfrak{G}_{D}^{(s)}$ for every state $s \in \mathcal{S}$, and thus there is no incentive for deviation ex-post.

Consider the ex-ante game $\mathfrak{G}_{D}\left(\mathcal{N},v_{n,D}\right)$ and let $\bar{\mathbf{x}}:=\bar{\mathbf{x}}_{\mathcal{N}}$ be the Walrasian equilibrium under uncertainty. Following the discussion above, we know that $\bar{\mathbf{x}}_{\mathcal{N}} \in \mathbb{C}
\left(\mathfrak{G}_{D}\right)$. Since $\bar{\mathbf{x}}_{\mathcal{N}}$ is a competitive equilibrium, $\bar{\mathbf{x}}_{n,\mathcal{N}}^{(s)}$, i.e., the reward allocated to agent $n \in \mathcal{N}$ at state $s \in \mathcal{S}$, maximizes $v_{n}$ on the budget set 
$\mathcal{B}_{n}(\mathbf{p})=\left \{\mathbf{x}_{n}:\mathbf{p}\cdot \bar{\mathbf{x}}_{n} \leq \mathbf{p} \cdot \mathbf{q}_{n} \right\}$. By 
(\ref{eq:DetEqivThree}), we know that $v_{n,D}\left(\left[\bar{\mathbf{x}}_{n}^{(s)}\right]_{s \in \mathcal{S}}\right)=\sum_{s=1}^{S}
a^{(s)}u_{n,D}^{(s)}\left(\bar{\mathbf{x}}_{n}^{(s)}\right)$. Moreover, $v_{n}$ is state-separable and $u_{n,D}^{(s)}$, $s \in \mathcal{S}$, are concave and monotonically increasing by Proposition \ref{pr:UtiAss}. Thus, $\bar{\mathbf{x}}_{\mathcal{N}}^{(s)}$ is the unique maximizer of $u_{n,D}^{(s)}$ on the budget set $\mathcal{B}_{n}^{(s)}(\mathbf{p})=\left \{\mathbf{x}_{n}^{(s)}:\mathbf{p}^{(s)}\cdot 
\bar{\mathbf{x}}_{n}^{(s)}\leq \mathbf{p}^{(s)}\cdot \mathbf{q}_{n,D}^{(s)}\right\}$. The rest of the proof is similar to proving that the competitive equilibrium lies inside the core \cite{Debreu63:ALT}.

Assume that at some state $s \in \mathcal{S}$, $\bar{\mathbf{x}}_{\mathcal{N}}^{(s)}$ does not belong to the core of the state game after the resolution of the uncertainty; i.e, $\bar{\mathbf{x}}_{\mathcal{N}}^{(s)} \notin \mathbb{C}\left(\mathfrak{G}_{D}^{(s)} \right)$. This implies that at state $s$, there is some coalition $c$ that can unilaterally improve upon $\bar{\mathbf{x}}^{(s)}_{\mathcal{N}}$ via an allocation $\mathbf{y}_{c}$. Formally,
\begin{equation}
\label{eq:axOne}
\sum_{n \in c} \mathbf{y}_{n,c}^{(s)} \leq \sum_{n \in c} \mathbf{q}_{n,D}^{(s)};
\end{equation}
\begin{equation}
\label{eq:axTwo}
\begin{matrix}
\forall ~ n\in c:& u_{n,D}^{(s)}\left(\mathbf{y}_{n,c}^{(s)} \right)\geq u_{n,D}^{(s)}\left(\bar{\mathbf{x}}_{n,\mathcal{N}}^{(s)} \right) ;
\end{matrix}
\end{equation}
\begin{equation}
\label{eq:axThree}
\begin{matrix}
\exists~ n\in c:& u_{n,D}^{(s)}\left(\mathbf{y}_{n,c}^{(s)} \right)>u_{n,D}^{(s)}\left(\bar{\mathbf{x}}_{n,\mathcal{N}}^{(s)} \right). 
\end{matrix}
\end{equation}
As argued before, $\mathbf{x}_{\mathcal{N}}^{(s)}$ maximizes $u_{n,D}^{(s)}$ on the budget set. Thus, (\ref{eq:axThree}) ensures that 
\begin{equation}
\label{eq:axFour}
\begin{matrix}
\exists~ n\in c:& \mathbf{p}^{(s)}\cdot \mathbf{y}_{n,c}^{(s)}>\mathbf{p}^{(s)}\cdot \mathbf{q}_{n,D}^{(s)}
\end{matrix}
\end{equation}  
Since if $\mathbf{p}^{(s)}\cdot \mathbf{y}_{n,c}^{(s)}<\mathbf{p}^{(s)}\cdot \mathbf{q}_{n,D}^{(s)}$, there would be a neighborhood $A$ of 
$\mathbf{y}_{n,c}^{(s)}$ for which there exists some $\mathbf{x}_{n,c} \in A$ so that $\mathbf{p}^{(s)}\cdot \mathbf{x}_{n,c}^{(s)}< 
\mathbf{p}^{(s)}\cdot \mathbf{q}_{n,D}^{(s)}$, and by the locally non-satiated (LNS) property, or due to monotonically increasing assumption of utility (see Proposition \ref{pr:UtiAss}), such a neighborhood contains an $\mathbf{x}_{n,c}^{(s)}$ that satisfies $u_{n,D}^{(s)}
\left(\mathbf{x}_{n,c}^{(s)}\right)>u_{n,D}^{(s)}\left(\mathbf{y}_{n,c}^{(s)} \right)\geq u_{n,D}^{(s)}\left(\bar{\mathbf{x}}_{n,\mathcal{N}}^
{(s)} \right)$, which is inconsistent with $\bar{\mathbf{x}}_{n,\mathcal{N}}^{(s)}$ maximizing $u_{n,D}^{(s)}$ on the budget set. Then 
(\ref{eq:axTwo}) yields
\begin{equation}
\label{eq:axFive}
\begin{matrix}
\forall~ n\in c:& \mathbf{p}^{(s)}\cdot \mathbf{y}_{n,c}^{(s)}> \mathbf{p}^{(s)}\cdot \mathbf{q}_{n,D}^{(s)}
\end{matrix}
\end{equation}  
By summing the inequalities (\ref{eq:axFour}) and (\ref{eq:axFive}) over coalition $c$, we have
\begin{equation}
\label{eq:axSix}
\begin{aligned}
\sum_{n\in c} \mathbf{p}^{(s)}\cdot \mathbf{y}_{n,c}^{(s)}&>\sum_{n\in c}\mathbf{p}^{(s)}\cdot \mathbf{q}_{n,D}^{(s)}& \\ 
\Rightarrow \mathbf{p}^{(s)}\cdot \sum_{n\in c}& \mathbf{y}_{n,c}^{(s)}> \mathbf{p}^{(s)}\cdot \sum_{n\in c} \mathbf{q}_{n,D}^{(s)}&
\end{aligned}
\end{equation}  
Since $\mathbf{p}^{(s)} \in \mathcal{R}_{\geq 0}$, we have $\sum_{n\in c} \mathbf{y}_{n,c}^{(s)}>\sum_{n\in c}\mathbf{q}_{n,D}^{(s)}$ for at least one $n \in c$, which contradicts (\ref{eq:axOne}).
\subsection{Proof of Proposition \ref{pr:WalConver}}
\label{subSec:ThCon}
As described in the proof of Proposition \ref{Pr:WalChar}, the utility function given by (\ref{eq:DetEqivThree}) satisfies the gross substitutes property. Then by Lemma \ref{Lm:ExcDem}, the excess demand function satisfies the gross substitutes property as well. Therefore, the result follows by Theorem \ref{Th:Conv}. 

\bibliographystyle{IEEEbib}
\bibliography{references}
\end{document}